\documentclass[twocolumn]{aastex631}

\newcommand{\gal}{\mbox{Galactocentric }}
\newcommand{\rg}{\mbox{$R_{G}$}}
                 
\newcommand{\msun}{\mbox{M$_\odot$}}               
\newcommand{\hii}{\mbox{\ion{H}{2}}}

\shorttitle{Disk Fraction in the Outer Milky Way}
\shortauthors{Patra et al.}
\graphicspath{{./}{figures/}}

\begin{document}

\title{Does Metallicity affect Protoplanetary Disk Fraction? \\Answers from the Outer Milky Way }

\correspondingauthor{Sudeshna Patra \& Jessy Jose}
\email{inspire.sudeshna@gmail.com, jessyvjose1@gmail.com}

\author[0000-0002-3577-6488]{Sudeshna Patra}
\affiliation{Indian Institute of Science Education and Research (IISER) Tirupati, Rami Reddy Nagar, Karakambadi Road, Mangalam(P.O.), Tirupati 517507, India}

\author[0000-0003-4908-4404]{Jessy Jose}
\affiliation{Indian Institute of Science Education and Research (IISER) Tirupati, Rami Reddy Nagar, Karakambadi Road, Mangalam(P.O.), Tirupati 517507, India}

\author[0000-0001-5175-1777]{Neal J. Evans II}
\affiliation{Department of Astronomy The University of Texas at Austin, 2515 Speedway, Stop C1400 Austin, Texas 78712-1205, USA}



\begin{abstract}
The role of metallicity in shaping protoplanetary disk evolution remains poorly comprehended.
This study analyzes the disk fraction of 10 young (0.9-2.1 Myr) and low-metallicity (0.34-0.83 Z$_{\odot}$) clusters located in the outer Milky Way with Galactocentric distances between 10 and 13 kpc.
Using $JHK$ data obtained from UKIDSS, the calculated disk fraction values for low-mass stars (0.2-2 \msun) ranged from 42\% to 7\%.
To enhance the statistical reliability of our analysis, eight additional low-metallicity clusters are sourced from previous studies with metallicity range 0.25-0.85 Z$_{\odot}$
along with our sample, resulting in a total of 18 regions with low-metallicity. 
We find that low-metallicity clusters exhibit on average $2.6\pm0.2$ times lower disk fraction compared to solar-metallicity clusters in all the age bins we have.
Within the age range we can probe, our study does not find evidence of faster disk decay in sub-solar metallicity regions compared to solar-metallicity regions.
Furthermore, we observe a positive correlation between cluster disk fraction and metallicity for two different age groups of $0.3-1.4$ and $1.4-2.5$ Myr.
We emphasize that both cluster age and metallicity significantly affect the fraction of stars with evidence of inner disks.
\end{abstract}



\section{Introduction} \label{sec:intro}
The protoplanetary disks (PPDs) around young stars, filled with gas and dust, act as progenitors of planetary systems 
by supplying the raw material for planet formation \citep{2003ARA&A..41...57L}. 
The lifetime of the protoplanetary disks is important in determining the time frame 
for planet formation and the physical processes that lead to its dissipation \citep{2011ARA&A..49...67W}. 

Measuring the frequency of stars with infrared excess in a sample of clusters is one of the most common methods to study the disk lifetime 
\citep{2001ApJ...553L.153H, 2003ARA&A..41...57L}.
Disks emit strong radiation at various wavelengths ranging from microns to millimeters due to the temperature gradient.
The innermost portion of the disk, consisting of hot dust, primarily contributes to near-infrared (NIR) continuum emission, while cooler regions farther from the star emit at longer wavelengths \citep{1998AREPS..26...53B}.
Emission in excess of the photosphere in the near-infrared ($< 10 \mu m$) is a strong indicator of PPDs \citep{2011ARA&A..49...67W}.
The presence of NIR excess is well correlated with spectroscopic signatures of accretion, 
which allows for the investigation of the inner accretion disk lifetimes (radius $<\sim$ 0.1 AU) 
through the study of the fraction of stars exhibiting this excess as a function of their age \citep{1995ApJ...452..736H}. 

The dissipation timescales of PPDs may vary between the inner and outer regions of the disk \citep{2009ApJ...705.1237G, 2021PASJ...73.1589M}.
However, the studies by \cite{2005ApJ...631.1134A} and \cite{2014MNRAS.442.2543Y} suggest that for low-mass stars, the entire disk disperses almost simultaneously. 
The infrared studies \citep{2007ApJ...662.1067H, 2007prpl.conf..573M, 2009AIPC.1158....3M, 2018MNRAS.477.5191R}  have shown that the majority of disks tend to disappear within a few million years. For solar neighborhood clusters, the disk fraction falls exponentially with cluster age. Young clusters have a high disk fraction of $60\%$ to $80\%$, which drops to around $20\%$ after 5 Myr \citep{2001ApJ...553L.153H,2006ApJ...638..897S,2014A&A...561A..54R, 2015A&A...576A..52R}.
Recently, with a more complete data sets down to low-mass stars, \citet{2022ApJ...939L..10P} has shown that for nearby low-mass stars, the e-folding timescale for disk dissipation is $\sim$ 7 Myr.

The primary mechanisms responsible for disk dissipation are (i) mass accretion onto the host star, (ii) the dissipation due to photoevaporation, and/or (iii) planet formation. The interplay of these processes ultimately determines the extent of the disk dispersal \citep{2011ARA&A..49...67W}.   
There are more environmental feedback mechanisms (e.g., external-photoevaporation, stellar encounter) that can influence the disk dissipation process significantly, but these mostly affect the outer disk.
In the external-photoevaporation process, a nearby massive OB-type star irradiates the disk with high-energy photons such as far-ultraviolet (FUV) and extreme ultraviolet (EUV) \citep{1999ApJ...515..669S,2001MNRAS.325..449S,2016MNRAS.457.3593F,2016arXiv160501773G}. 
In high-density parental clusters, close encounters between members can lead to mutual gravitational interactions that disperse disk material into the medium or capture it in other stars \citep{1993MNRAS.261..190C,2005A&A...437..967P,2005MNRAS.364..961T,2015MNRAS.446.2010M}. 

Another important but less explored factor is the metallicity of disks, which is also anticipated to have a significant role in determining disk dispersion timescales through its influence on dust content \citep{2010ApJ...723L.113Y}. 
Exoplanet surveys highlight the impact of disk metallicity on the occurrence rate of planets \citep{2005ApJ...622.1102F, 2012A&A...541A..97M, 2018AJ....156..221N, 2018AJ....155...89P}, especially for giant planets \citep{2021ApJS..255...14F}.

Disks with low-metallicity have higher ionization fraction, leading to higher accretion rates driven by magnetorotational instability, which is more efficient with increasing disk ionization \citep{2006ApJ...648..484H}. 
Theory predicts that both mass accretion and mass loss due to photoevaporation, are expected to increase significantly with increasing penetration depth of ultraviolet photons at low metallicity \citep{2016ApJ...818..152B}. In particular, far-ultraviolet (FUV) radiation penetrates deeper into disks with low dust opacity, thus decreasing the disk dispersal timescale \citep{2009ApJ...690.1539G}. 
\citet{2010MNRAS.402.2735E, 2018ApJ...857...57N, 2018ApJ...865...75N} suggest that low metallicity increases the efficiency of photoevaporation in removing the gas and dust grains from the disks. 
The disk dispersal timescale due to photoevaporation increases with the disk metallicity.

NIR-based observational studies by \cite{2009ApJ...705...54Y, 2010ApJ...723L.113Y, 2016AJ....151..115Y,2016AJ....151...50Y, 2021AJ....161..139Y} have found that the rate of disk dispersal is much faster in low-metallicity environments compared to solar metallicity. For low-metallicity clusters studied by \cite{2009ApJ...705...54Y, 2010ApJ...723L.113Y, 2016AJ....151..115Y,2016AJ....151...50Y, 2021AJ....161..139Y}, 
the disk fraction is $\lesssim20\%$ even at 1 Myr and $\sim5\%$ after 2 Myr.
Recently, \cite{2021A&A...650A.157G} also suggested that metallicity may be more important than other environmental processes in causing the faster disk dissipation rate for the young cluster Dolidze 25 (metallicity $-0.5$ dex below solar for oxygen).
The observed trend that disk lifetimes increase with metallicity is consistent with the result from theoretical simulations \citep{2010MNRAS.402.2735E, 2018ApJ...857...57N, 2018ApJ...865...75N}. 
This highlights the importance of considering metallicity in understanding the evolution of disks around young stars. 
However, \citet{2023ApJS..267...46I} did not find the impact of metallicity on accretion luminosities for a sub-solar metallicity ($\sim0.6-0.7\ Z_{\odot}$) region CMa$-l224$ in the Outer Galaxy. 
Also, there are contradictory results found by \cite{2010ApJ...715....1D, 2011ApJ...740...11D, 2017ApJ...846..110D, 2023A&A...675A.204V} where they reported that the mass-accretion process is longer in low metallicity star-forming complexes in the Large and Small Magellanic Clouds (LMC, SMC). They measured directly the infall rate of the gas onto the star using H$\alpha$ based photometric method instead of the apparent dust content of circumstellar disks. 
Their proposed explanation for the prolonged duration of the accretion process in low-metallicity environments is that the lower radiation pressure exerted by the forming star on the low-metallicity disc material results in less efficient disc dispersion \citep{2017ApJ...846..110D}. 

The impact of metallicity on the protoplanetary disk evolution remains poorly understood due to a scarcity of observational tests. 
The Milky Way has a negative metallicity gradient with increasing Galactocentric distance, making the outer part of the Milky Way a laboratory of low-metallicity environment \citep{2000MNRAS.311..329D, 2017MNRAS.471..987E,2018MNRAS.478.2315E, 2020MNRAS.496.1051A,2022MNRAS.510.4436M, 2023NatAs...7..951L}.

We more than double the sample and perform a homogeneous analysis to answer whether metallicity affects disk dissipation timescales.
In this paper, we analyze NIR data from 10 low-metallicity clusters located in the outer Milky Way and supplement our findings with data from 8 low-metallicity clusters studied previously by  \cite{2009ApJ...705...54Y,2010ApJ...723L.113Y,2016AJ....151..115Y, 2016AJ....151...50Y, 2021AJ....161..139Y}.
Finally, we compare the disk fraction between low and solar metallicity regions.
Following this introduction, Section \ref{sec:sample} describes the sample and data used in this paper. Section \ref{sec:results} explains the method used for selecting candidate members in clusters (section \ref{subsec:all_members}), followed by estimation of age and extinction (section \ref{subsec:age}) and disk fraction (section \ref{subsec:diskfrac}). 
In section \ref{sec:discussion}, we discuss the results from previous studies on low-metallicity and solar metallicity regions (section \ref{subsec:previous}), discuss the dependency of disk fraction on both cluster age and metallicity (section \ref{subsec:df_vs_Z_age}) 
and the relevance of metallicity over other factors (section \ref{subsec:other_factors}).
The study concludes by acknowledging the limitations of this study in section \ref{subsec:caveats}, and provides a summary of the findings in section \ref{sec:conclusion}.

\section{Target and Data} \label{sec:sample}
\begin{figure*}[ht!]
\epsscale{0.9}
\plotone{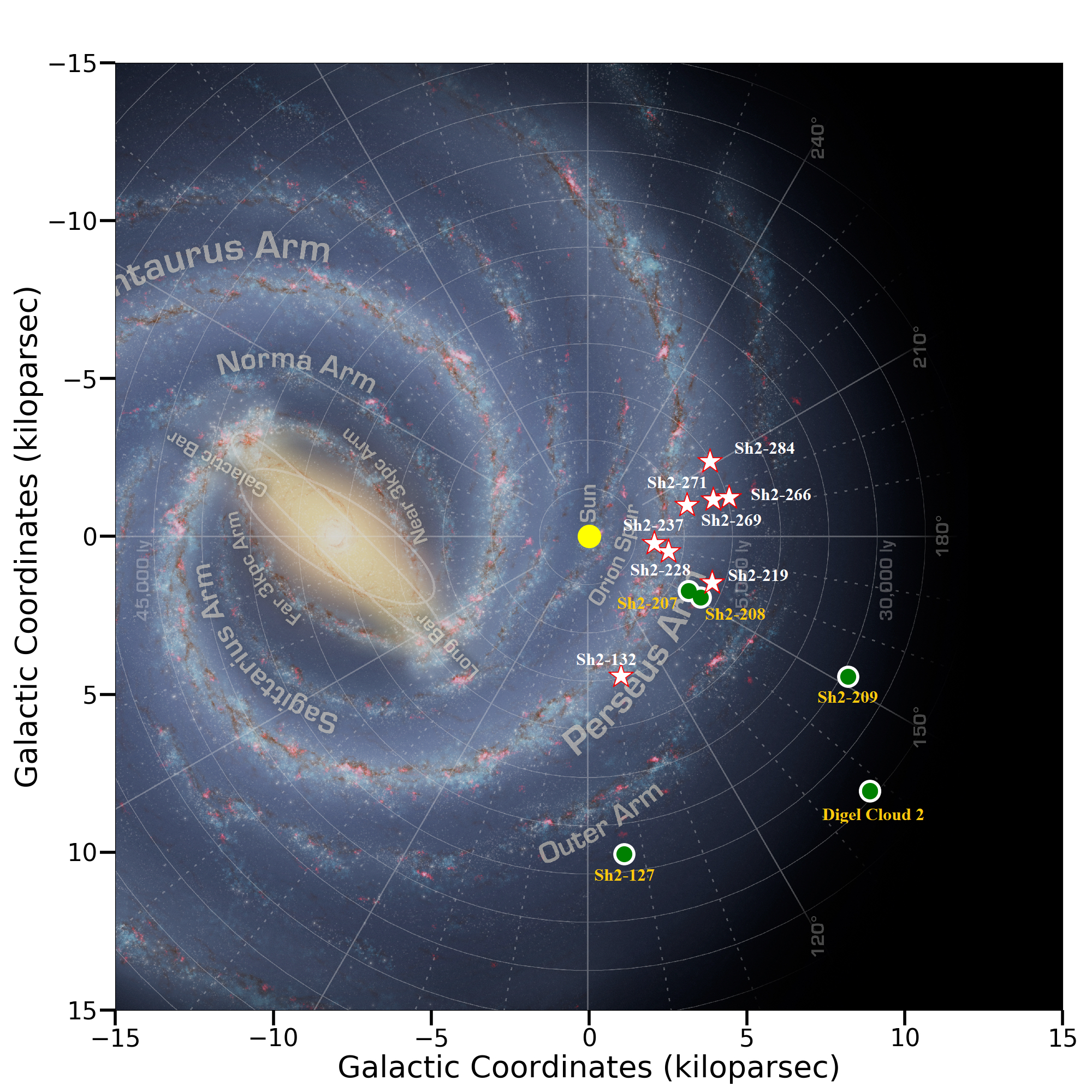}
\caption{Spatial distribution of 10 young clusters listed in Table \ref{tab:cloud details} are shown with white star marks and the previously studied 8 clusters (from \citealt{2009ApJ...705...54Y, 2010ApJ...723L.113Y, 2016AJ....151..115Y,2016AJ....151...50Y, 2021AJ....161..139Y}) are shown with green filled circles. 
\label{fig:target}}
\end{figure*}

We have selected 10 young star clusters with a primary emphasis on their \gal\ distance (\rg), availability of deep NIR data, and well-determined metallicity.
The clusters included in this study for comprehensive statistical analysis, namely Sh2-132, Sh2-219, Sh2-228, Sh2-237, Sh2-266, Sh2-269, and three sub-clusters in Sh2-284, are associated with \hii \ regions.
All of the selected targets are located beyond the solar circle ($\rg>8.2$ kpc, \citealt{2019A&A...625L..10G}) in the second and third quadrants, and are indicated by white star markers in Figure \ref{fig:target} relative to both Sun and Galactic center.
The \rg\ range of all the targets falls within $10-13$ kpc.
We have sourced distance information for all targets, except Sh2-269 and Sh2-284, from \citet{2022MNRAS.510.4436M}, where they have derived these values using GAIA EDR3 data. 
For Sh2-269 and Sh2-284, the distance information are taken from \citet{2019A&A...625A..70Q,2011MNRAS.410..227C}, respectively. 
We have derived the heliocentric distances ($D$) of the targets from their corresponding \rg, and the distances range from $2-4.5$ kpc.
The \gal\ (\rg) and heliocentric distance ($D$) information for all the targets are provided in Table \ref{tab:cloud details}. 
We have used 12+log[O/H] values from \citet{2022MNRAS.510.4436M} for all targets except Sh2-228, Sh2-269, and Sh2-284. For Sh2-228 and Sh2-269, we have obtained the values from \citet{2018PASP..130k4301W}, and for Sh2-284, we have used the value from \citet{2015A&A...584A..77N}.
We have converted the 12+log[O/H] values of the targets to the metallicity ($Z$) which is the metallicity relative to the solar neighborhood ($Z=Z_{*}/Z_{\odot}=10^{[M/H]}$), where $Z_{\odot}=0.0143$ \citep{2009ARA&A..47..481A}. The [M/H] is the difference between the 12+log[O/H] value of the target and solar neighborhood ISM. We considered 12+log[O/H]=8.50 for solar neighborhood ISM based on \cite{2022ApJ...931...92E}. 
The $Z$ range for the selected regions is $0.34-0.85 \ Z_{\odot}$. 
We have computed the uncertainty of relative metallicity, which is derived from the propagated uncertainty originating from the 12+log[O/H] value of the target and the solar neighborhood ISM.
The individual relative metallicity values and the corresponding uncertainties are mentioned in Table \ref{tab:disk_fraction}.

\begin{deluxetable*}{l c c l l c l c c }[h]
    \tablenum{1}
    \tabletypesize{\footnotesize}
    \tablecaption{Target Details \label{tab:cloud details}}
    \tablewidth{0pt}
    \tablehead{
    \colhead{Target} & \colhead{RA}    & \colhead{DEC}   &  \colhead{\rg \ [Ref]}  & \colhead{$D$}    & \colhead{ Radius} & \colhead{Radius}  & \colhead{$J$ band Completeness} & \colhead{$A_{V}$}\\
    \colhead{}    & \colhead{(deg)} & \colhead{(deg)} & \colhead{(kpc)} & \colhead{(kpc)}  & \colhead{(arcmin)} & \colhead{(pc)}   & \colhead{(mag)} & \colhead{(mag)}
    }
    \startdata
        Sh2-132       &  $334.7529$    & $+56.1071$  & $10.21^{+0.28}_{-0.26}$ [1]   & $4.53^{+0.29}_{-0.27}$ &   2.0   & 2.63   & 19.5      &   $2.6 \pm 0.9$ \\
        Sh2-219       &  $74.0173$     & $+47.3778$  & $12.17^{+0.39}_{-0.41}$ [1]   & $4.16^{+0.32}_{-0.28}$ &   1.1   & 1.30   & 19.5      &   $4.3 \pm 1.1$ \\
        Sh2-228       &  $78.3662$     & $+37.4392$  & $10.72^{+0.19}_{-0.19}$ [1]   & $2.56^{+0.09}_{-0.09}$ &   2.0   & 1.50   & 18.5      &   $3.5 \pm 0.9$ \\
        Sh2-237       &  $82.8524$     & $+34.2123$  & $10.26^{+0.16}_{-0.17}$ [1]   & $2.07^{+0.06}_{-0.06}$ &   3.3   & 1.98   & 18.5      &   $2.1 \pm 1.0$ \\
        Sh2-266       &  $94.6878$     & $+15.2791$  & $12.69^{+0.62}_{-0.65}$ [1]   & $4.60^{+0.53}_{-0.51}$ &   2.0   & 2.67   & 20.0      &   $3.6 \pm 1.1$ \\
        Sh2-269       &  $93.6514$     & $+13.8248$  & $12.57^{+0.25}_{-0.21}$ [2]   & $4.06^{+0.37}_{-0.31}$ &   1.5   & 1.77   & 19.5      &   $4.0 \pm 2.5$ \\
        Sh2-271       &  $93.7332$     & $+12.3465$  & $11.34^{+0.22}_{-0.22}$ [1]   & $3.25^{+0.11}_{-0.12}$ &   1.3   & 1.23   & 19.5      &   $3.4 \pm 1.4$ \\
        Sh2-284 (C-1) &  $101.2581$    & $+00.2230$   & $11.70^{+0.16}_{-0.15}$ [3]  & $4.50^{+0.30}_{-0.30}$ &   1.2   & 1.57   & 19.0      &   $2.8 \pm 1.1$ \\
        Sh2-284 (C-2) &  $101.1914$    & $+00.3282$   & $11.70^{+0.16}_{-0.15}$ [3]  & $4.50^{+0.30}_{-0.30}$ &   1.7   & 2.16   & 19.0      &   $2.4 \pm 1.3$ \\
        Sh2-284 (C-3) &  $101.5992$    & $-00.0562$   & $11.70^{+0.16}_{-0.15}$ [3]  & $4.50^{+0.30}_{-0.30}$ &   2.0   & 2.62   & 19.0      &   $3.4 \pm 1.4$ \\
        \hline
    \enddata
    \tablecomments{[1] \citet{2022MNRAS.510.4436M}, [2] Quiroga-Nu{\~n}ez et al. (2019), [3] \citet{2011MNRAS.410..227C} \\
     The parameters without references are derived in this work. See section \ref{sec:sample}, \ref{subsec:all_members} and \ref{subsec:age} in the main text.}
\end{deluxetable*}

%
For each cluster, we have used the photometry in $J$, $H$, and $K$-bands from the Galactic Plane Survey (GPS) of the United Kingdom Infrared Deep Sky Survey (UKIDSS; \citealt{2007MNRAS.379.1599L}) DR11 plus database, which was observed using the WFCAM (Wide Field Camera) mounted on the 3.8-m United Kingdom Infrared Telescope (UKIRT).
We applied $p(star)>0.9$ to ensure the source is a star, not a galaxy or noise, and $PriOrSec(m)=0$ to exclude the repeating sources in the overlapped regions \citep{2008MNRAS.391..136L}. 
We considered the sources fainter than 13 mag in the $J$ band because UKIDSS photometry has a saturation limit at  $J \sim13$ mag \citep{2008MNRAS.391..136L} and also with photometric uncertainty less than $0.2$ mag in all the bands. 
The fainter end of the photometric data is usually incomplete due to several factors, including observational sensitivity, crowding, and variable extinction. To assess the completeness of the photometry used in our analysis for each cluster, we plot histograms of the sources detected within the analysis area (discuss in section \ref{subsec:all_members}). The faintest magnitude bin having more than $50\%$  of the peak value in the histogram is considered as the $50\%$  data completeness limit for a cluster.
We consider the magnitude corresponding to the 50\% completeness bin as the lower magnitude limit for all the targets and the corresponding $J$-band magnitudes are mentioned in Table \ref{tab:cloud details}.
We have also selected a nebulosity-free region as the control field, located within the vicinity of $\sim$  $3-10\arcmin$ distance from the cluster field for each target (see more details in section \ref{subsec:all_members}).

\section{Analysis and Results} \label{sec:results}
To estimate the disk fraction, the two main steps are identifying the probable candidate members in the cluster and identifying the members with disks among them. Finally, to understand the disk evolution, estimating the age of the cluster is another important step. 
The following sections explain the details of each step.

\subsection{Candidate members in the cluster} \label{subsec:all_members}
\paragraph{Cluster radius}
The first step is to define the cluster radius to find the member stars of the cluster. We compute the stellar surface density and identify the region with higher density compared to the surroundings as the cluster area.
We use the most reliable and well-known approach to obtain the stellar surface density, the k-nearest neighborhood method \citep{1985ApJ...298...80C}. 
The generalized form of the $j^{th}$ nearest neighbor surface density for a star is $(\rho_{j}=(j-1)/\pi r_{j}^{2})$, where $r_{j}$ is the distance from one particular star to its $j^{th}$ neighbor. 
The choice of appropriate $j$ value is a non-trivial task since opting for small values of $j$ will provide spurious sub-clustering, whereas higher $j$ values will lead to the detection of large-scale structures \citep{2021MNRAS.504.2557D}. For our analysis, we choose an optimal value of  $j=15$ to estimate the cluster radius.
%
Next, we calculated the background counts, mean ($\mu$), and standard deviation ($\sigma$) of the stellar density distribution within the control field. 
The control field has the same radius as the cluster field, and the field star distribution in the control field is expected to be similar to that in the cluster field due to their proximity.
We consider the highest density point as the center of the cluster, while the cluster radius is determined by finding the density level that corresponds to $3\sigma$ contour above the mean ($\mu$) density observed in the control field.
For more details, see \citet{2021MNRAS.504.2557D}. The corresponding cluster radii are listed in Table \ref{tab:cloud details}.

\begin{figure*}[ht!]
\epsscale{1}
\plotone{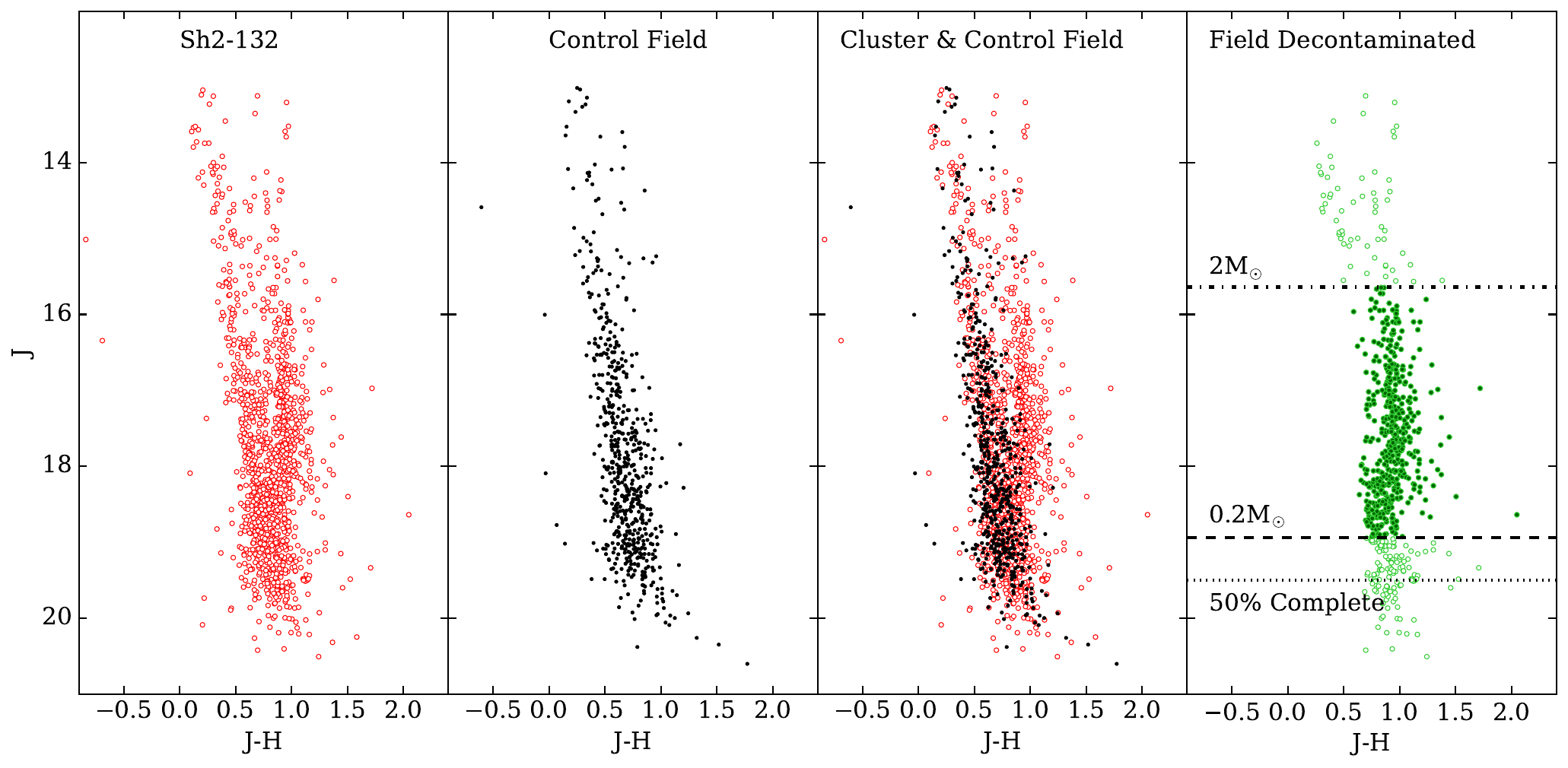}
\caption{\textit{Panel 1:} $J-H$ vs $J$ CMD of all the sources within the radius of  Sh2-132. 
\textit{Panel 2:} CMD for the nearby control field of a similar area. 
\textit{Panel 3:} Cluster and control field CMDs are overplotted. 
\textit{Panel 4:} CMD of the cluster after statistically subtracting the field stars. Green open circles denote the probable members of the cluster. Green-filled circles correspond to the finally selected members for the analysis after giving the mass limits in the lower and upper end (see text for details).  The respective mass limits and the photometric completeness limits are marked in the figure.\label{fig:field_decon}}
\end{figure*}

\paragraph{Field Star Decontamination}
Spectroscopic observations or multi-band SED analysis are the best ways to identify the membership in young clusters (e.g. \citealt{2017MNRAS.468.2684P, 2020ApJ...892..122J} and references therein).
The clusters in this study are relatively distant and not all UKIDSS sources have counterparts in other wavelengths. Hence, these membership analysis methods are beyond this paper's scope.
We follow the statistical field subtraction method to obtain the probable number of cluster members \citep{2017ApJ...836...98J,2020ApJ...896...29K, 2021MNRAS.508.3388G}.
The region within the cluster radius includes the cluster members along with the foreground and background population in that direction. 
We have statistically subtracted the nearby control field from the cluster region to separate out the probable members from the field population using their $(J-H)$ vs $J$ distribution (see \citealt{2021MNRAS.504.2557D} for details). 
Figure \ref{fig:field_decon} illustrates the $(J-H)$ vs $J$ color-magnitude diagrams (CMDs) for a sample cluster Sh2-132 (first panel) and its associated control field (second panel). 
 In the CMD of the cluster, there are two sequences - the sequence appearing to the right is mainly contributed by the young, pre-main sequence members of the cluster, whereas, the sequence to the left is contributed by the field population towards the direction of the cluster. The sequence to the right is absent in the CMD of the control field (panel 2). 
The third panel of Figure \ref{fig:field_decon} represents the  CMDs of cluster and control fields overplotted, and it shows that the field sequence at the left of both cluster field and control fields are overlapping on each other.  
Typically, the young star clusters associated with \hii\ regions are likely to have more extinction than the control field due to the presence of additional dust associated with the underlying molecular cloud. 
However, in Figure \ref{fig:field_decon},  the  field sequences of both the cluster and control fields overlie, suggesting a negligible extinction difference between them. This ensures we can perform statistical subtraction without correcting for extra extinction in the control field.

We have followed the method by \cite{2021MNRAS.504.2557D} to subtract the field stars from the cluster field 
where we divided the colour and magnitude axes of $(J-H)$ vs $J$ CMD for both the cluster and the control field regions into bins of size 0.1 and 0.2 mag, respectively.
We have subtracted the number of sources in each bin of the control field from the number of sources in the corresponding bin of the cluster region.
In order to improve the decontamination process and enhance the accuracy of our final catalog, we use a Gaussian function to define the locus of the PMS branch.
We fit the function along the color axis for bins of 0.5 magnitude in the $J$ band. Since the PMS branch is nearly vertical, we fit the Gaussian perpendicular to the distribution.
The peak of the Gaussian curve determines the mean locus of the PMS branch.
We eliminated the scattered sources located beyond the $1\sigma$ limit of the mean locus of the pre-main sequence (PMS) branch, but only on the bluer side of the CMD. 
We did not impose the $1\sigma$ restriction on the redder side of the mean locus of PMS since sources with NIR excess or relatively high reddening are likely to be present there.
Panel 4 of Figure \ref{fig:field_decon} represents the probable members after statistical subtraction, which essentially retains the sequence due to the young stellar objects at the right. 
The respective CMDs for other clusters analyzed in this study are provided in Appendix \ref{appen:A}.

\subsection{Extinction and Age estimation}\label{subsec:age}
Understanding a star cluster's evolution and formation history relies significantly on the age and age spread of its constituent stars \citep{2003ARA&A..41...57L}. Estimating the accurate age information of young clusters is challenging due to inherent uncertainties, particularly during their early stages of formation \citep{2014prpl.conf..219S}. 
We often simplify the scenario by assuming that all stars in a given cluster formed simultaneously from the interstellar medium.  However, numerous studies have demonstrated non-coeval stellar evolution in young star-forming regions, leading to spread in the age estimation (e.g., \citealt{2016ApJ...822...49J, 2017ApJ...838..150K, 2018AJ....155...44P, 2023ApJ...948....7D} and others).
The standard method followed for age estimation is the use of the Hertzsprung-Russell (HR) diagram to find the positions of candidate members on the locations of theoretical PMS evolutionary tracks and isochrones \citep{2015ApJ...808...23H}. 

For age estimation, we need to remove the extinction associated with the individual clusters. 
We consider the field decontaminated candidate PMS stars (see Section \ref{subsec:all_members}) for extinction estimation of each cluster. 
We have assumed the interstellar reddening law of $A_{\rm J}/A_{\rm V}=0.243$; $A_{\rm H}/A_{\rm V} = 0.131$ and $A_{\rm K}/A_{\rm V} = 0.078$ from \citet{2019ApJ...877..116W} to derive the K-band extinction (i.e., $A_{\rm K}=E(J-H)\times0.697$, where $E(J-H)=(J-H)_{obs}-(J-H)_{int}$).  
The intrinsic color $(J-H)_{int}$ is taken as 0.6 mag, which is the median value of $(J-H)$ colors of K and M type stars \citep{2013ApJS..208....9P}. 
We fit a Gaussian function for each cluster on the histogram distribution of $A_{\rm K}$ of all the candidate PMS sources. 
The peak of the Gaussian function is taken as the mean extinction of the cluster, while the standard deviation of the function is considered as the spread in the extinction associated with the cluster. The mean $A_{\rm V}$ values of our sample vary from $2.1-4.0$ mag with an uncertainty range of $0.9-2.5$ mag. The extinction values for each cluster are mentioned in Table \ref{tab:cloud details}.

We obtained the HR diagram by estimating the luminosity and temperature of a limited sample of field-subtracted PMS members in each cluster. We achieved this by performing Spectral Energy Distribution (SED) analysis using the online tool VO Sed Analyzer (VOSA)\footnote{http://svo2.cab.inta-csic.es/theory/vosa/}. 
This is an online tool which makes SEDs by sourcing additional wavelength information from diverse online photometric catalogues (such as 2MASS, Spitzer, GLIMPSE, WISE, SDSS, GAIA EDR3, IPHAS, Pan-Starrs, VPHAS, wherever available) of the supplied $JHK$ data (see \citealt{2021MNRAS.504.2557D} for details). However, SED analysis could only be applied to a limited number of PMS stars in each cluster, as not all had counterparts at other wavelengths.
We also provided the distance, extinction, and spread in the extinction associated with each cluster (see Table \ref{tab:cloud details}) as the input in VOSA. VOSA uses this information to deredden and correct the observed SEDs. 
Subsequently, the SEDs are compared to synthetic photometry derived from the BT-Settl (AGSS2009) model \citep{2012IAUS..282..235A}. We use the models for respective metallicity for each cluster (see Table \ref{tab:disk_fraction}).
By performing a chi-square fitting analysis, VOSA estimates various stellar parameters, which include the temperature and luminosity of 
cluster members. 
Next, we use these luminosity and temperature values to estimate the age of the individual PMS sources by comparing their location on the HR diagram with the isochrones derived from the PARSEC  (PAdova and TRieste Stellar Evolution Code)\footnote{http://stev.oapd.inaf.it/cgi-bin/cmd} version 1.2S evolutionary models \citep{2012MNRAS.427..127B}. Since the clusters in this study are associated with \hii~ regions, we use isochrones of  age between 0.5-10 Myr with 0.1 Myr interval.  
We use the isochrones for the respective metallicity value of the clusters. 
The age of the closest isochrone is assigned as the age of a given source.  
The median and standard deviation values thus estimated from this method are considered to be the age and age spread of the individual clusters. 
However, estimating the age of clusters relies entirely on the chosen evolutionary models \citep{2014prpl.conf..219S, 2018MNRAS.477.5191R}.
The median age of the clusters in this study ranges from 0.9-2.1 Myr.
We have listed the median age of each cluster and their age spread in Table \ref{tab:disk_fraction}.

\subsection{Disk Fraction}\label{subsec:diskfrac}
The disk fraction, which represents the percentage of stars in a young stellar cluster that have dust disks, plays a significant role in studying the lifespan and development of these disks.
In this section, we calculate the NIR-excess-based disk fraction ($D_{f}$) of the 10 clusters mentioned in Table \ref{tab:cloud details} using the following equation. 
\begin{equation}\label{eq1}
    D_{f}\mathrm{[per\ cent] = \frac{N_{Disk}}{N_{Total}}} \times 100,
\end{equation}
where $N_{Total}$ represents the total number of probable member stars of each cluster and $N_{Disk}$ represents the number of stars in the observed sample that exhibit NIR excess due to the presence of inner dust disks.
To calculate the $N_{Total}$ for each cluster, we impose the mass limits in both the upper and lower ends of the distribution. 
Based on the distance, reddening, and age of the clusters, the mass limit corresponding to the indicated photometric completeness limit in Table \ref{tab:cloud details} varies from 0.05 to 0.2 \msun. 
The black dotted line in panel 4 of Figure \ref{fig:field_decon} and the figures in Appendix \ref{appen:A} indicate the 50\% photometric completeness of each target.  
To ensure consistency and uniformity across all targets, we have chosen a lower mass limit of 0.2 \msun\ across all the targets for analysis of disk frequencies. This mass limit is represented by the black dashed line in panel 4 of Figure \ref{fig:field_decon} and figures in Appendix \ref{appen:A}. 
We excluded stars with a mass greater than 2 \msun\ because  the stars with stellar mass $< 2$ \msun\ are generally considered to have similar disk evolution (\citealt{2015A&A...576A..52R, 2022ApJ...939L..10P}). 
Also, only $4-12\%$ stars have mass greater than 2 \msun\ in each cluster, so omitting these small numbers of stars would not significantly impact our disk fraction analysis.
The black dash-dotted line in panel 4 of Figure \ref{fig:field_decon} indicates the upper mass limit. Only the members that fall within these mass boundaries are considered as $N_{Total}$, and we have denoted them with green filled circles in panel 4 of Figure \ref{fig:field_decon} and also in Appendix \ref{appen:A}. 

\begin{figure}[ht!]
\epsscale{1.2}
\plotone{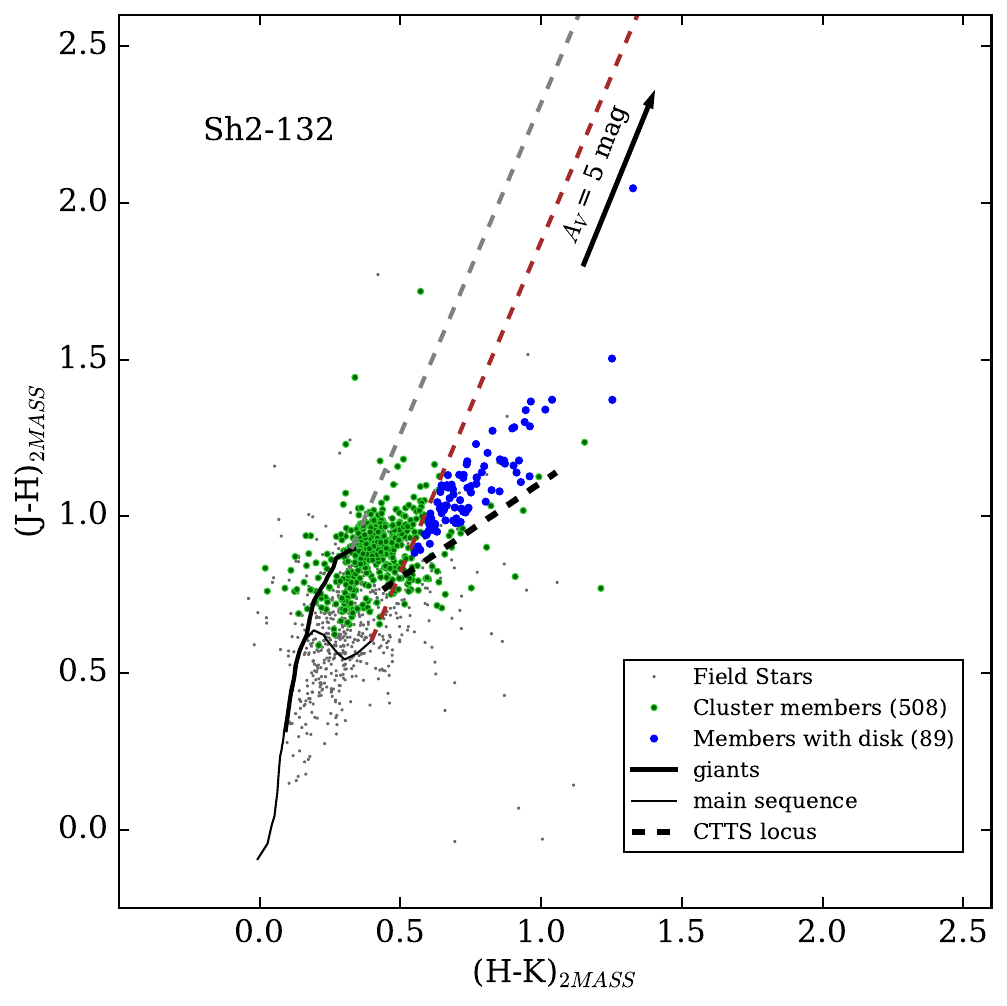}
\caption{$(J-H)$ vs. $(H-K)$ color–color diagrams for Sh2-132 cluster in the 2MASS system. The two solid curves in the lower left portion of the diagram represent the loci of the main sequence (thin line) and the giant stars (thicker line). 
The two parallel dashed lines are the reddening vectors.
The brown dashed line, which intersects the main-sequence curve at the maximum $H - K$ values (M6 spectral type) 
is the border between stars with and without circumstellar disks. The CTTS loci are shown by a black dashed line.  Identified cluster members and field stars are shown by green-filled circles and black dots, respectively. The stars with circumstellar disks are shown in blue dots. \label{fig:cc_diag}}
\end{figure}

To determine $N_{Disk}$, we have used the $J-H$ versus $H-K$ color-color (CC) diagram, which distinguishes stars with intrinsic excess emission, reddened stars due to interstellar dust, and stars with normal, unreddened photospheric colors \citep{1992ApJ...393..278L}. 
In Figure \ref{fig:cc_diag}, we have plotted the cluster members of Sh2-132 (green dots) and the field stars (black dots). We have also plotted the two black solid curves, which represent the loci of the main sequence (thin line) and the giant stars (thicker line) \citep{1988PASP..100.1134B}. 
The black dashed line is the classical T Tauri locus from \cite{1997AJ....114..288M}. 
Two parallel dashed lines constitute the reddening bands from the tip of the main sequence and giant star sequence following the above reddening law. 
The field stars, Class III sources without excess are located within these bands. 
The targets with intrinsic infrared excess emission populate the redward region of the brown dashed line and above the classical T Tauri locus \citep{2002ApJ...565L..25S,2004ApJ...608..797O,2004ApJ...616.1042O, 2015MNRAS.454.3597D}.
The colours and the curves shown in the figure are all transformed to the 2MASS photometric system. 
We used the equations from \cite{2001AJ....121.2851C} to do the color transformation between Bressell \& Brett (BB) homogenized system to 2MASS; Caltech (CIT) photometric system to 2MASS and UKIRT to 2MASS. 
In Figure \ref{fig:cc_diag}, some of the field stars are scattered and found at the location of NIR excess sources.  
Hence our NIR excess source list is also likely to have some field contamination.
In order to estimate the amount of possible contamination in the NIR excess source list, we reddened the control field by the additional reddening associated with the cluster. We then counted the number of field sources whose excess is more than  $1\sigma$ (where $\sigma$ is the mean color error) from the brown and black dashed lines in figure \ref{fig:cc_diag}.   
The amount of possible contaminants at the location of  NIR excess ranges between $3-18\%$ for the clusters in this study. 
We have considered only the sources above $1\sigma$ offset from the brown and black dashed lines as disk-bearing sources, which we have indicated as $N_{disk}$ and marked with blue dots in Figure \ref{fig:cc_diag}.

We use equation \ref{eq1} to compute the disk fraction (in percentage) for all the targets within the mass range of $2-0.2$ \msun. The resulting disk fraction values of 10 clusters range from 42\% to 7\%. 
Detailed information on the disk fraction and associated uncertainties of individual targets are provided in Table 2. The Poisson error is considered as the corresponding uncertainty in disk fraction estimation. 
The CC diagrams for the remaining targets are given in Appendix \ref{appen:B}, and we have mentioned the number of $N_{Total}$ and $N_{Disk}$ of each target in their respective CC diagrams.

\begin{deluxetable*}{l l c c c c}
    \tablenum{2}
    \tabletypesize{\footnotesize}
    \tablecaption{Disk fraction vs age \label{tab:disk_fraction}}
    \tablewidth{0pt}
    \tablehead{
    \colhead{Target}   &    \colhead{Metallicity}                  & \colhead{Age}           & \colhead{Disk Fraction}    & \colhead{Lower Mass Limit} & \colhead{References}\\
    \colhead{}         & \colhead{($Z=\frac{Z_{*}}{Z_{\odot}}$)}   & \colhead{(Myr)}         &   \colhead{(\%)}           & \colhead{($M_{\odot}$)}  &  \colhead{} }
    \startdata
        & \multicolumn{3}{c}{This work}\\
        \hline
        Sh2-132       &  0.74 (0.14) &         $1.7^{+3.7}_{-1.2}$     & 18$\pm$2   &   0.2     & [1]\\
        Sh2-219       &  0.60 (0.12) &         $1.3^{+2.2}_{-0.8}$     & 42$\pm$7   &   0.2     & [1]\\
        Sh2-228       &  0.78 (0.07) &         $1.5^{+3.5}_{-1.0}$     & 23$\pm$4   &   0.2     & [2]\\
        Sh2-237       &  0.69 (0.08) &         $2.1^{+3.5}_{-1.3}$     & 12$\pm$2   &   0.2     & [1]\\
        Sh2-266       &  0.49 (0.16) &         $2.1^{+4.9}_{-1.5}$     & 18$\pm$3   &   0.2     & [1]\\
        Sh2-269       &  0.83 (0.08) &         $0.9^{+2.1}_{-0.6}$     & 39$\pm$8   &   0.2     & [2]\\
        Sh2-271       &  0.59 (0.09) &         $2.1^{+4.0}_{-1.4}$     & 23$\pm$4   &   0.2     & [1]\\
        Sh2-284 (C-1) &  0.34 (\nodata) &      $1.3^{+2.9}_{-0.9}$     & 7$\pm$3    &   0.2     & [3]\\
        Sh2-284 (C-2) &  0.34 (\nodata) &      $1.7^{+2.6}_{-1.0}$     & 7$\pm$2    &   0.2     & [3]\\
        Sh2-284 (C-3) &  0.34 (\nodata) &      $2.1^{+3.9}_{-1.4}$     & 11$\pm$3   &   0.2     & [3]\\        
        \hline
        & \multicolumn{3}{c}{Low metallicity clusters from \citet{2009ApJ...705...54Y, 2010ApJ...723L.113Y, 2016AJ....151..115Y, 2016AJ....151...50Y, 2021AJ....161..139Y}}\\
        \hline
        Sh2-127A      &   0.57 (0.04)    & 0.5     [$0.9^{+0.7}_{-0.4}$]$^{*}$   & 28$\pm$3     &   0.2           & [1]  \\
        Sh2-127B      &   0.57 (0.04)    & 0.1-0.5 [$0.9^{+0.7}_{-0.4}$]$^{*}$    & 40$\pm$4     &    0.2         & [1] \\
        Sh2-207       &   0.66 (0.07)    & 2-3     [$1.7^{+3.7}_{-1.2}$]$^{*}$    &  4$\pm$2     &    0.07-0.08   & [2] \\    
        Sh2-208       &   0.85 (0.07)    & 0.5                                    & 27$\pm$6     &    0.05        & [2] \\
        Sh2-209 (Main)&   0.38 (0.26)    & 0.5-1   [$1.3^{+2.2}_{-0.8}$]$^{*}$    & 10$\pm$0.8   &    0.08-0.09   & [1] \\
        Sh2-209 (Sub) &   0.38 (0.26)    & 0.5-1   [$1.3^{+2.2}_{-0.8}$]$^{*}$    & 7.1$\pm$1.2  &    0.08-0.09   & [1] \\
        Cloud 2-N     &   0.25 (\nodata)    & 0.5-1                               &  9$\pm$4     &    0.06        & [4] \\
        Cloud 2-S     &   0.25 (\nodata)    & 0.5-1                               & 27$\pm$7     &    0.06        & [4] \\
        \hline
        & \multicolumn{3}{c}{Nearby solar metallicity clusters from literature ($JHK$)} \\
        & \multicolumn{3}{c}{(references there in \citealt{2009ApJ...705...54Y, 2010ApJ...723L.113Y})}\\
        \hline
        Orion OB1b    &    0.99 (0.27) &   5.0      &  8.9$\pm$3     &   0.10           &    $\dagger$    \\
        Upper Sco     &    1.00 (0.28) &   5.0      &  32$\pm$4      &   0.10           &    $\dagger$    \\
        $\eta$ Cham   &    1.00 (0.28) &   6.0      &  28$\pm$12     &    0.08          &    $\dagger$    \\
        Orion OB1a    &    0.99 (0.27) &   8.5      &  3.6$\pm$2.5   &   0.10           &    $\dagger$     \\
        NGC 7160      &    0.99 (0.27) &   10.0     &  6.3$\pm$3.6   &   0.40           &    $\dagger$    \\
        NGC 2024      &    0.95 (0.26) &   0.3      &  58$\pm$7      &   0.13           &    $\dagger$    \\
        Trapezium     &    0.95 (0.26) &   1.5      &  53$\pm$3      &   $\sim$ 0.03    &    $\dagger$    \\
        Cham I        &    1.00 (0.28) &   1.8      &  41$\pm$4      &   $\sim$ 0.03    &    $\dagger$    \\
        Taurus        &    0.98 (0.27) &   2.3      &  34$\pm$6      &   0.30           &    $\dagger$    \\
        IC 348        &    0.97 (0.26) &   2.3      &  21$\pm$4      &   0.19           &    $\dagger$    \\
        NGC 2264      &    0.98 (0.27) &   3.3      &  30$\pm$3      &   $\sim$ 0.03    &    $\dagger$    \\
        Tr 37         &    0.99 (0.27) &   4.0      &  17$\pm$3      &   $\sim$ 0.03    &    $\dagger$    \\
        \hline
    \enddata
    \tablecomments{[1] \citet{2022MNRAS.510.4436M} [2] \citet{2018PASP..130k4301W}; [3] Negueruela et al. (2015); [4] \citet{2010ApJ...723L.113Y} \\
    $^{*}$Age in square bracket is obtained using our method. \\
    $\dagger$ The metallicity is calculated by using radial gradient equation from \cite{2022MNRAS.510.4436M} for the nearby clusters.}
\end{deluxetable*}

\section{Discussion} \label{sec:discussion}

\subsection{Combining the Data-set with Previous Studies} \label{subsec:previous}
We aim to enhance the statistical robustness of our low-metallicity data-set by incorporating data from prior investigations on low-metallicity samples conducted by \cite{2009ApJ...705...54Y, 2010ApJ...723L.113Y, 2016AJ....151..115Y, 2016AJ....151...50Y, 2021AJ....161..139Y}. This inclusion will contribute to a more comprehensive and nuanced understanding of the effect of metallicity within our current study.

\cite{2009ApJ...705...54Y, 2010ApJ...723L.113Y, 2016AJ....151..115Y,2016AJ....151...50Y, 2021AJ....161..139Y} studied total of 8 clusters - one cluster each in the \hii~ regions Sh2-207 \citep{2016AJ....151...50Y}  and Sh2-208 \citep{2016AJ....151..115Y}, and two clusters each in Sh2-209 \citep{2010ApJ...723L.113Y}, Sh2-127 \citep{2021AJ....161..139Y}, and Digel Cloud 2 \citep{2009ApJ...705...54Y} [hereafter ``Yasui’s sample”]. The spatial distribution of these targets can be seen in Figure \ref{fig:target} (green filled circles), and they are located within the \rg\ range of 11.5-19 kpc. 
These targets are observed using the $JHK$-bands of MOIRCS (Multi-Object InfraRed Camera and Spectrograph) of 8-m Subaru telescope, and they probe down to lower stellar masses (0.05-0.2 \msun) compared to our mass limit of $0.2\ \msun$. In Table \ref{tab:disk_fraction}, we have listed the disk fraction, age, and mass limit values of these clusters (Sh2-127(A,B), Sh2-207, Sh2-208, Sh2-209 (main and sub-clusters), Cloud 2-N and S) from the literature. 
Their disk fraction measurements are also based on NIR-excess analysis; however there are slight variations in the specifics.

In Yasui's sample, since the reddening in the clusters was relatively high and hence they considered the sources lying towards the right side of the CMD, which are highly reddened, as cluster members and neglected the contamination from field stars.
For the identification of sources with disks, they employed the $J-H$ versus $H-K$ color-color diagrams in MKO system and considered all sources positioned to the right of the reddening vector corresponding to an M6 star (which serves as the boundary between sources with and without disks). 
To calculate the disk fraction, they assumed disk emission is primarily observable in the $K$ band.
They measured the age by conducting a comparison between observed and model KLFs (K-band Luminosity Functions). 
In order to cross-check the age estimation, we applied similar method as in this work (explained in Section \ref{subsec:age}) to calculate the ages of the clusters that were reported by \cite{2010ApJ...723L.113Y, 2016AJ....151...50Y, 2021AJ....161..139Y}  for which UKIDSS data are available. Table \ref{tab:disk_fraction} also includes the resulting age values for these clusters. 
We generally got older ages for those clusters than their literature values except Sh2-207; however, the deviations are not significant. Considering the uncertainties involved, the age estimations from both methods agree for most of the clusters.

We have also considered the disk fraction and age information of 12 embedded clusters with solar metallicity in the solar neighborhood, as presented in \citealt{2009ApJ...705...54Y} (Table 1) and \citealt{2010ApJ...723L.113Y} (Table 2). Among these clusters, seven are characterized as young clusters (with ages $\leqslant5$ Myr, including NGC 2024, Trapezium, Cham I, Taurus, IC 348, NGC 2264, Tr 37), while the remaining five clusters exhibit relatively older ages ($\sim5-10$ Myr, encompassing Orion OB1b, Upper Sco, $\eta$ Cham, Orion OB1a, NGC 7160). 
The disk fraction in these ``local" clusters provides a comparison sample for the low metallicity clusters.
The lower stellar mass limit of these solar metallicity clusters varies from $0.03-0.4$ \msun\ and the individual values are mentioned in Table \ref{tab:disk_fraction}.
The authors have obtained the $ JHK$-based disk fraction of these clusters in a similar way as low-metallicity clusters and also measured age from the KLF analysis using photometry data from the literature. 
Since the precise metallicity information was not available for this sample, we have calculated their metallicity by using the $R_{G}$ versus metallicity relation from \cite{2022MNRAS.510.4436M},  where the 
distance is taken from \citet{2022ApJ...939L..10P}.
All relevant information has been compiled in Table \ref{tab:disk_fraction}.

\begin{figure*}[t]
\epsscale{1.2}
\plotone{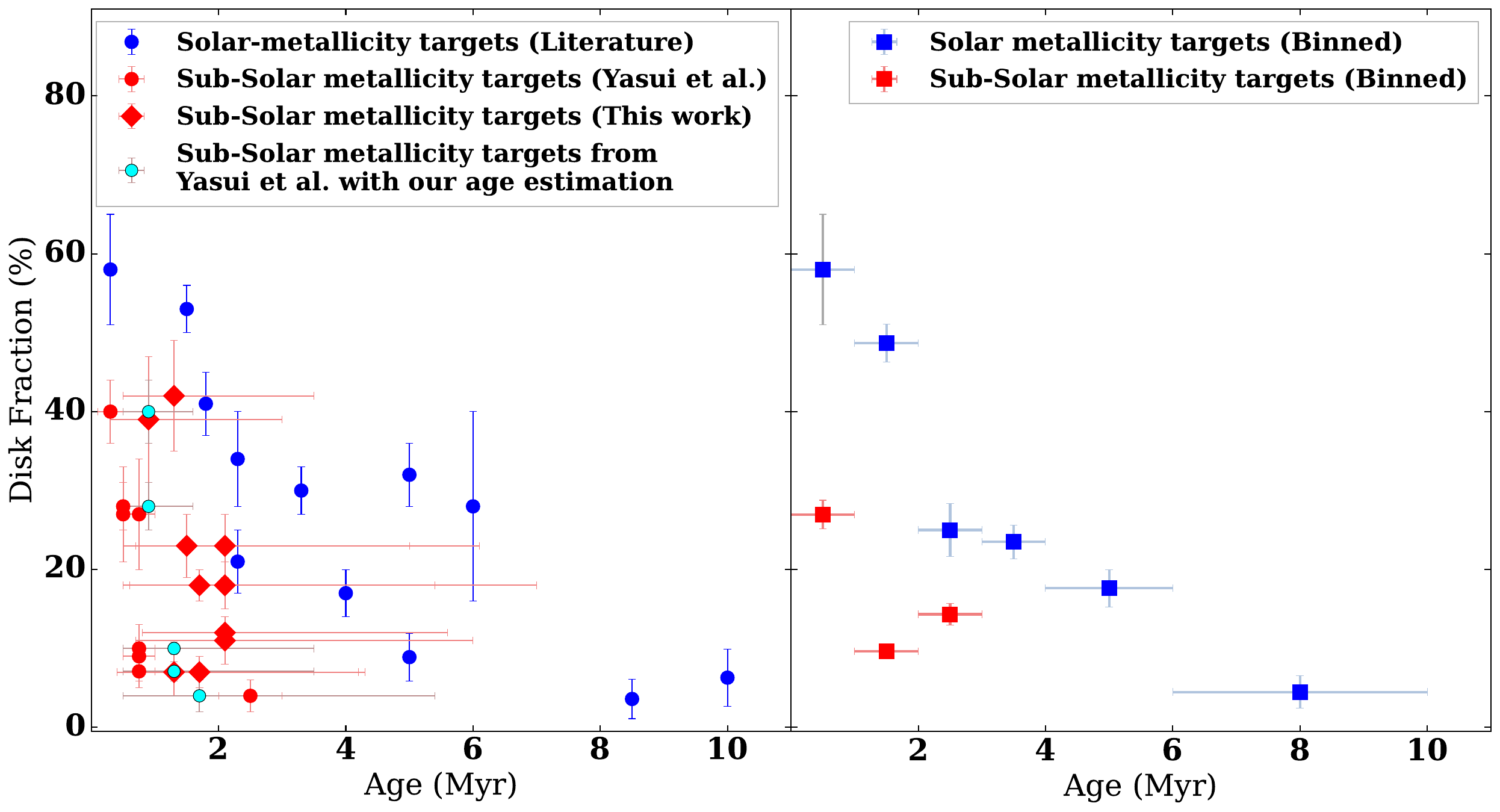}
\caption{\textit{Left:} Disk fraction ($D_{f}$) vs cluster age ($t$) distribution based on JHK data for the 10 low-metallicity clusters of this work (red diamonds), Yasui’s sample of low-metallicity (8 clusters, red filled circles) and solar metallicity regions (12 clusters, blue filled circles). The cyan filled circles are for Yasui’s clusters (Sh2-127 A, B; Sh2-207 and the main and sub-clusters of Sh2-209) with our age estimation.
\textit{Right:} Disk fraction ($D_{f}$) vs cluster age ($t$) distribution for binned data. The solar metallicity clusters (blue squares) are  binned in 1 Myr age span upto 4 Myr and next two points show 2 Myr age binning. All low metallicity clusters are binned in 1 Myr age span up to 3 Myr, denoted by red squares.
\label{fig:df_vs_age}}
\end{figure*}

The left panel of Figure \ref{fig:df_vs_age} illustrates a comparison between regions of solar and sub-solar metallicity. The red diamonds represent sub-solar targets from this study, the red filled circles denote sub-solar targets adopted from Yasui's sample. 
The cyan filled circles in the left panel of Figure \ref{fig:df_vs_age} portray the disk fraction values sourced from \citep{2021AJ....161..139Y,2016AJ....151...50Y, 2010ApJ...723L.113Y}, while incorporating our own age estimations for the targets Sh2-127 A, B; Sh2-207 and the main and sub clusters of Sh2-209.
The 12 blue points correspond to solar metallicity targets sourced from the literature \citep{2009ApJ...705...54Y, 2010ApJ...723L.113Y}.

\cite{2021A&A...650A.157G} has recently investigated the timescale for the dispersal of protoplanetary disks in the Dolidze 25 region, which is associated with the \hii\ region Sh2-284. Dolidze 25 is one of the well-known low-metallicity young clusters located in the outer Galaxy.
Using a multiwavelength catalog combining optical, infrared, and X-ray data, they found a disk fraction of $\sim 34\%$ and a median age of 1.2 Myr for the whole complex. Our sample includes the sub-clusters S284-C1 and S284-C2 of the Dolidze 25 complex. 
The disk fraction value we obtained is much lower than the disk fraction values reported by \cite{2021A&A...650A.157G}. 
In the case of S284-C1, our analysis identified a total of 100 stars as potential member candidates, with 7 of them exhibiting evidence of disks. Conversely, \cite{2021A&A...650A.157G},  reported a count of 55 stars in total for the same region,  of which 12 displayed disk signatures. Similarly, for S284-C2, we detected 180 probable members, with 13 showing near-infrared excess indicative of disks. In contrast, \cite{2021A&A...650A.157G} identified 43 stars and found 28 candidates displaying disk features.
One possible reason for this discrepancy is - the study by \cite{2021A&A...650A.157G} is mainly limited by the sensitivity and resolution of \textit{Spitzer}-IRAC observations and hence the fainter, non-excess sources are mostly undetected in their sample. However, in our sample, the fainter end is mostly complete because of the better spatial resolution and sensitivity of UKIDSS.
Another possible reason is - they have detected more stars with disks than our $JHK$-based analysis because those sources are likely to be showing excess in longer wavelengths but not in shorter wavelengths. This is in agreement with \citet{2014A&A...561A..54R}, that the disk fraction depends on the wavelengths used for measurements and it increases systematically if measured at longer wavelengths.

The heterogeneity of disk fraction estimates across the literature is widely acknowledged. In this paper, the disk fraction of the samples under discussion are all based on near-infrared (NIR) excess analysis, which predominantly targets the inner portion of the disk. Consequently, we have incorporated low-metallicity samples from Yasui's dataset and compared them with Solar neighborhood samples, all derived from NIR excess studies. Although there are minor differences in the specifics, we anticipate minimal deviation in the values, given the consistent focus on the inner disk region across the analyzed samples.

\begin{figure*}[t]
\epsscale{0.7}
\plotone{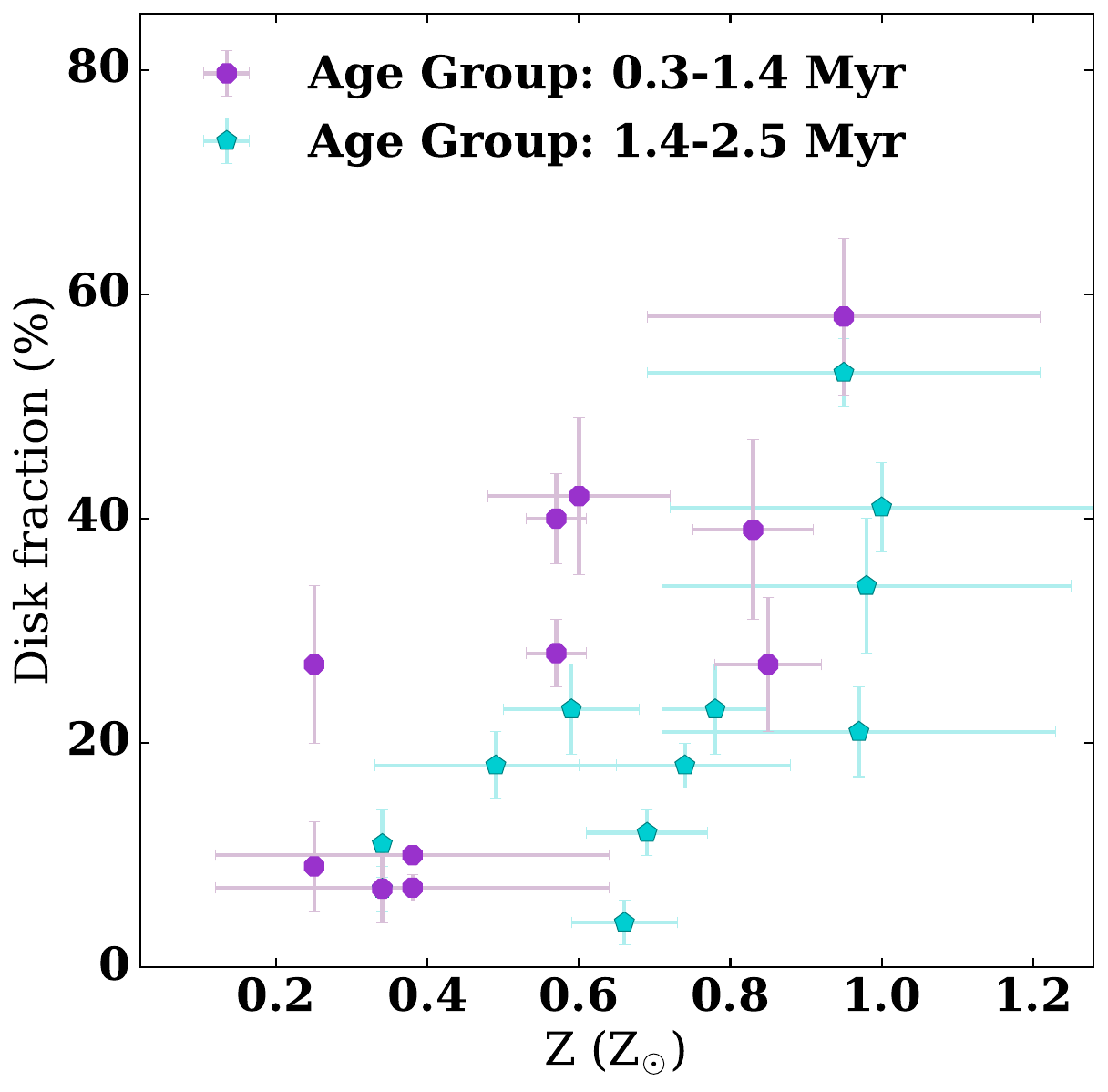}
\caption{Disk fraction $(D_{f})$ vs metallicity $(Z)$ plot based on $JHK$ data for two different age ranges.  
 Purple octagons represent the targets within the age range of 0.3 to 1.4 Myr and cyan pentagons represent targets within the age range of 1.4 to 2.5 Myr. 
\label{fig:df_vs_Z}}
\end{figure*}


\subsection{Disk fraction as a function of Cluster Age and Metallicity} \label{subsec:df_vs_Z_age}
The studies by \citet{2001ApJ...553L.153H, 2006ApJ...638..897S, 2007ApJ...662.1067H, 2007prpl.conf..573M, 2009AIPC.1158....3M, 2014A&A...561A..54R, 2015A&A...576A..52R, 2016ApJ...822...49J, 2018MNRAS.477.5191R, 2023JApA...44...77D} 
showed that disk fraction varies with age, while \citet{2009ApJ...705...54Y, 2010ApJ...723L.113Y, 2021AJ....161..139Y} have proposed that the disk fraction is influenced by metallicity.

\subsubsection{Disk Fraction versus Cluster Age}
The left panel of Figure \ref{fig:df_vs_age} shows that the disk fractions in sub-solar metallicity clusters generally lie significantly below those in solar metallicity clusters. Because both age and $Z$ clearly affect disk fraction, we binned the clusters in age to allow comparison of clusters with similar, but not identical ages.
First, we categorized all low-metallicity targets (mentioned in Table \ref{tab:disk_fraction}) into three age bins, each spanning 1 Myr. 
For the targets Sh2-127 A, B; Sh2-207 and Sh2-209 main and sub clusters, we have considered the age values obtained from our own age estimation method for consistency.
Similarly, we grouped solar-metallicity regions into age bins with a 1 Myr span, except for the last 2 age bins. 
We have calculated the weighted mean and errors in the weighted mean for each age bin, accounting for the uncertainties in disk fraction estimates.
In the right panel of Figure \ref{fig:df_vs_age}, the red and blue squares represent the weighted mean of the disk fraction values within each age bin for sub-solar and solar metallicity regions, respectively. 
The vertical error bars indicate the errors in the weighted mean of
the disk fraction within each age bin, while the horizontal error bars depict the age bin spread. 
For the first bin of solar metallicity regions, only one target is available and the vertical error bar denotes the individual uncertainty of that target.
%
The right panel of Figure \ref{fig:df_vs_age} illustrates a compelling comparison between solar and sub-solar metallicity regions, revealing a 
trend of lower disk fractions in sub-solar metallicity regions across all age bins. 
The ratio between the average disk fraction values of solar and sub-solar regions are $2.1\pm0.3$, $5.0\pm0.4$, and $1.7\pm0.3$ for age bins 0-1 Myr, 1-2 Myr, and 2-3 Myr, respectively. 
%
The observed pattern shows that the disk fraction in regions with lower metallicity is on average, $2.6\pm0.2$ times lower than that in solar metallicity regions. 
In Appendix \ref{appen:c}, we present a comparable plot (see Figure \ref{fig:df_vs_age_patra}), considering only 10 sub-solar clusters of our study, denoted by the red points. 
This independent analysis reinforces the conclusion that the disk fraction in low-metallicity clusters is lower when compared to clusters of solar metallicity, 
with a factor of $2.2\pm0.2$ on average across all age bins. 
This conclusion holds true even without considering Yasui's sample. 
Our result agrees with the findings from \cite{2010ApJ...723L.113Y, 2021AJ....161..139Y}, indicating a lower disk fraction in a low-metallicity environment compared to solar metallicity targets.
 The lower disk fraction ratios has two plausible explanations. Firstly, the initial disk fraction in low-metallicity regions is inherently lower compared to regions with solar metallicity.  Secondly, a significant proportion of disks in low-metallicity environments dissipate within a time frame shorter than our sample can measure.
Because we do not observe any systematic decline in the disk fraction ratios within the three age bins (Figure \ref{fig:df_vs_age}), such an early dissipation would have to occur in less than 1 Myr, the age of clusters in our youngest bin.

\subsubsection{Disk Fraction versus Metallicity}
To better understand the impact of metallicity, we examined the relationship between disk fraction and metallicity while keeping the age range of the sample constant. We divided the targets (both low and solar metallicity regions) into two groups with age ranges of 0.3-1.4 Myr and 1.4-2.5 Myr, for an even distribution of targets in each age bin, as shown in Figure \ref{fig:df_vs_Z}. 
In 
Figure 5, the purple octagons denote targets within the age group of 0.3-1.4 Myr 
and the cyan pentagons represent targets in the age range 1.4-2.5 Myr.
The vertical and horizontal error bars correspond to the uncertainties in disk fraction and metallicity estimations, as detailed in Table \ref{tab:disk_fraction}, for each target.

The data presented in Figure \ref{fig:df_vs_Z} demonstrate a positive correlation between disk fraction and metallicity across two age bins.
We have checked the statistical significance of the correlation coefficient by calculating the Pearson correlation coefficient (r-value) and the p-value for both the age groups. 
For the age range 0.3-1.4 Myr, the $r$ and $p-$values estimated are 0.77 and 0.005, respectively, while for the 1.4-2.5 Myr age group, the values are 0.72 and 0.007, respectively. 
The $r-$values indicate a strong positive correlation between disk fraction and metallicity for both the age groups and the associated p-values indicate that this correlation is statistically significant. 
These findings support the notion that disk fraction increases with rising metallicity across different age groups.

From Figure \ref{fig:df_vs_age} and \ref{fig:df_vs_Z}, we find that the disk fraction varies both with cluster age and metallicity. Specifically, the disk fraction tends to decrease with increasing cluster age, while showing an increasing trend with metallicity.


\subsection{Metallicity versus Other Factors} \label{subsec:other_factors}
Protoplanetary disk evolution depends on the environment in which they were born \citep{2007A&A...462..245G, 2014ApJ...794..147P}. In addition to metallicity, factors such as cluster density and the proximity of massive stars also have an impact on disk evolution. \cite{2018MNRAS.478.2700W} suggests that in dense stellar environments with a local stellar density ($n_{c}$) greater than $\mathrm{10^4\ pc^{-3}}$, close encounters between PPDs and other members increase the probability of collisions. 
These collisions significantly shape the distribution of PPD radii over a 3 Myr period by causing truncation and mass loss, affecting their overall size and distribution. 
Thus, considering local stellar density is important when studying the formation and evolution of PPDs in diverse environments.
External photoevaporation is another key mechanism in the evolution of planet-forming disks (see \citealt{2022EPJP..137.1132W} for more details). It occurs when the high FUV and EUV luminosities of an OB star heat up the outer material of the disk. The heating of the outer disk material, which experiences weaker gravity, combined with the intense radiation, leads to an extremely rapid outside-in depletion of the disk. 
Both of these effects are more prominent in the outer part of the disk, but it can affect the inner disk evolution as well (\citealt{2020RSOS....701271P, 2022EPJP..137.1132W} and references therein).

To see if external irradiation or the dynamical interactions affect the disk evolution in our sample, we calculated the local stellar density and the FUV flux emitted from the massive stars in our clusters. 
S132, S219, and S271 exemplify the most extreme cases in our sample, potentially more affected by FUV radiation due to the presence of O-type stars. Each of these clusters contains one massive O-type star, resulting in a corresponding FUV flux of $\sim 10^{3}G_{0}$ \citep{2016arXiv160501773G} and the local stellar density associated with them is ($n_{c}$) $\mathrm{\sim 10\ pc^{-3}}$. 
However, even in these cases, the FUV fields are significantly weaker than those experienced in the core of Orion Nebula Cluster (ONC). These clusters align with the lower end of ONC when positioned in $n_{c}-G_{0}$ space (\citealt{2018MNRAS.478.2700W}, Figure 3).
Additionally, when compared to Figure 15 of \cite{2018MNRAS.478.2700W}, our clusters are not heavily affected by either external photoevaporation or tidal truncation.
Thus, we can conclude that rapid disk dissipation in our clusters is not caused by these external effects, and instead suggest that metallicity is the main differentiating factor. \cite{2021A&A...650A.157G} also suggested the same for Dolidze 25, which is part of the Sh2-284 star-forming complex.

\subsection{Caveats and Future Work} \label{subsec:caveats}
Our analysis has two main caveats: (i) limited range of ages in the sample; and (ii) mass limitation in the sample.
The low-metallicity regions we examined had a relatively narrow age range of 0.5 to 3 Myr, compared to solar regions with a wider age coverage ranging from 0.3 to 10 Myr. 
Despite the age restriction, we observe that the disk fraction for low-metallicity clouds is lower than that for solar-metallicity clouds in the same age bin.
Incorporating low-metallicity targets with older ages ($\sim5-10$ Myr) will help in understanding the influence of metallicity on the disk evolution during the later stages.

We have restricted our analysis by imposing the lower mass limit of 0.2 \msun\ to maintain uniformity throughout the sample.
We have also calculated the disk fraction for each target by considering all the targets above the 50\% photometric completeness limit, thus including further low-mass stars ($<$ 0.2 \msun). 
The difference between the disk percentage values including low-massive stars, i.e, $<$ 0.2 \msun\ above 50\% completeness for the sample varies within a range of $-3.3\%$ to $1.5\%$, indicating small differences.
Highly sensitive observations, which are possible with the James Webb Space Telescope (JWST), should be able to capture a complete mass function, 
thus enabling in-depth exploration of the disk fraction down to the brown dwarf population for even more distant regions. 
This capability will serve as a robust foundation for various studies focusing on the formation of stars and planets with low metallicity. 
Moreover, by observing further distant and low-metallicity clusters in the outer Galaxy with JWST, we can achieve higher spatial resolution, which in turn can enhance the quality of our studies.

\section{Conclusions} \label{sec:conclusion}
In this paper, we study 10 low metallicity clusters in the outer Galaxy using the $JHK$ data from UKIDSS. We have calculated the age and disk fraction of these clusters. The main objective of this paper is to understand the role of metallicity on the disk fraction using a statistically robust sample. We have incorporated the result of 8 low-metallicity clusters from \cite{2009ApJ...705...54Y, 2010ApJ...723L.113Y, 2016AJ....151..115Y,2016AJ....151...50Y, 2021AJ....161..139Y}.
The results are summarized below.
\begin{enumerate}
    \item The fraction of disks is influenced by both the age and metallicity of the clusters.
    
    \item The disk fraction in the low-metallicity regions is on average $2.6\pm0.2$ times lower than solar metallicity regions.
    Our study does not provide definitive evidence of more rapid disk decay in sub-solar metallicity targets within the age range analyzed in this paper.

    \item The cluster disk fraction correlates with metallicity for both younger and older age bins.
    The Pearson correlation coefficient values ($r$ and $p$) for both the age group 
    of 0.3-1.4 Myr and 1.4-2.5 Myr
    denote a positive correlation between the metallicity and disk fraction.

    \item The clusters in this study are not located in extreme environments, so the external photoevaporation and stellar encounter processes are not effective here in reducing the disk fraction rapidly. 
    This fact highlights the significance of metallicity in determining the disk fraction.
    
\end{enumerate}


\begin{acknowledgments}
The authors acknowledge the referee for the constructive feedback, which have contributed to improving the clarity of the paper.
SP thanks DST-INSPIRE Fellowship (No. IF180092) of the Department of Science and Technology to support the Ph.D. program. 
JJ acknowledges the financial support received through the DST-SERB grant SPG/2021/003850. NJE thanks the Department of Astronomy at the University of Texas at Austin for ongoing research support.
We use the data from the UKIRT Infrared Deep Sky Survey. 
We acknowledge the use of VOSA, a tool developed as part of the Spanish Virtual Observatory project supported by the Spanish MINECO through grant AyA2017-84089, and it has been partially updated with support from the European Union's Horizon 2020 Research and Innovation Programme, under Grant Agreement n$^{\circ}$ 776403 (EXOPLANETS-A). 
We would like to thank Drisya Karinkuzhi for the discussion on the metallicity of the regions, Manoj Puravankara and Manash Samal for their valuable comments, which helped to improve the paper. We thank Belinda Damian and Pratibha Sazawal for the useful discussion on various analyses.
\end{acknowledgments}

%

\vspace{5mm}
\facilities{UKIDSS}


\software{astropy\citep{2013A&A...558A..33A,2018AJ....156..123A,2022ApJ...935..167A},  
          NumPy \citep{2011CSE....13b..22V,2020Natur.585..357H},
          matplotlib \citep{2007CSE.....9...90H},
          IPython \citep{2007CSE.....9c..21P},
          SciPy \citep{2022zndo....595738G}.}



\clearpage
\appendix
\section{Appendix A}\label{appen:A}
\begin{figure}[ht!]
\epsscale{1}
\plotone{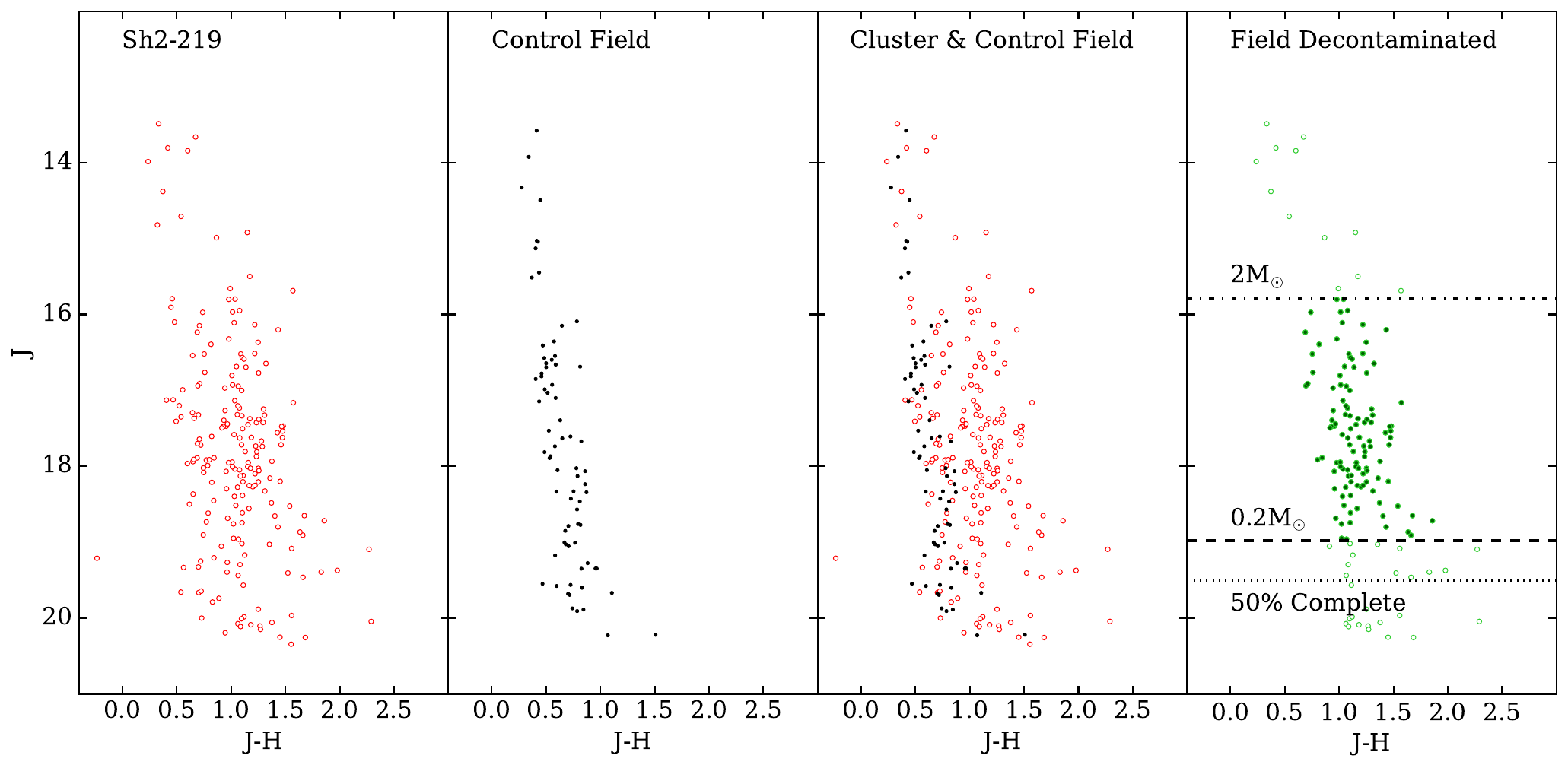}
\caption{CMDs of Sh2-219 \label{Appen_a1}}
\end{figure}

\begin{figure}[ht!]
\epsscale{1}
\plotone{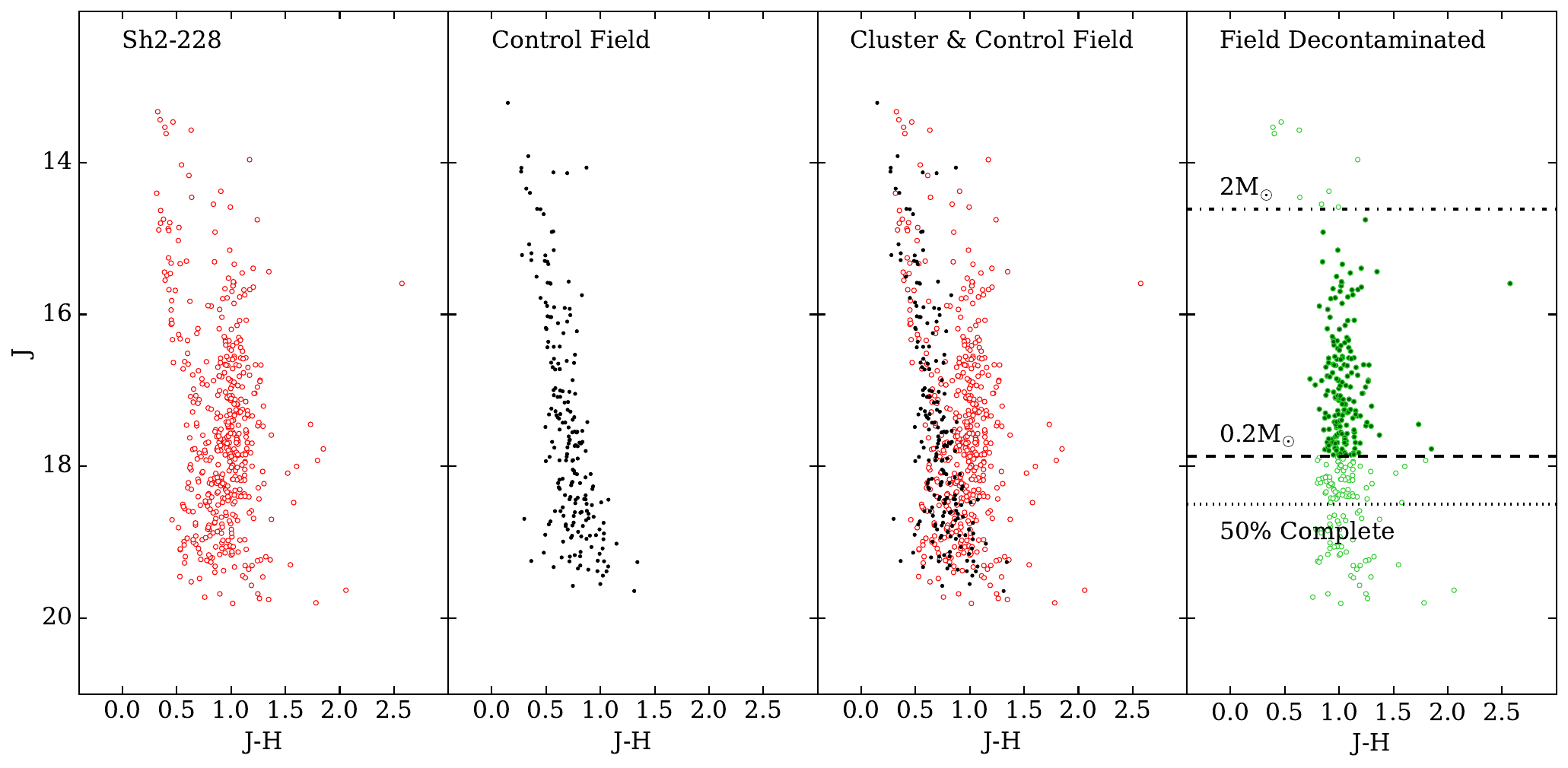}
\caption{CMDs of Sh2-228 \label{Appen_a2}}
\end{figure}

\begin{figure}[ht!]
\epsscale{1}
\plotone{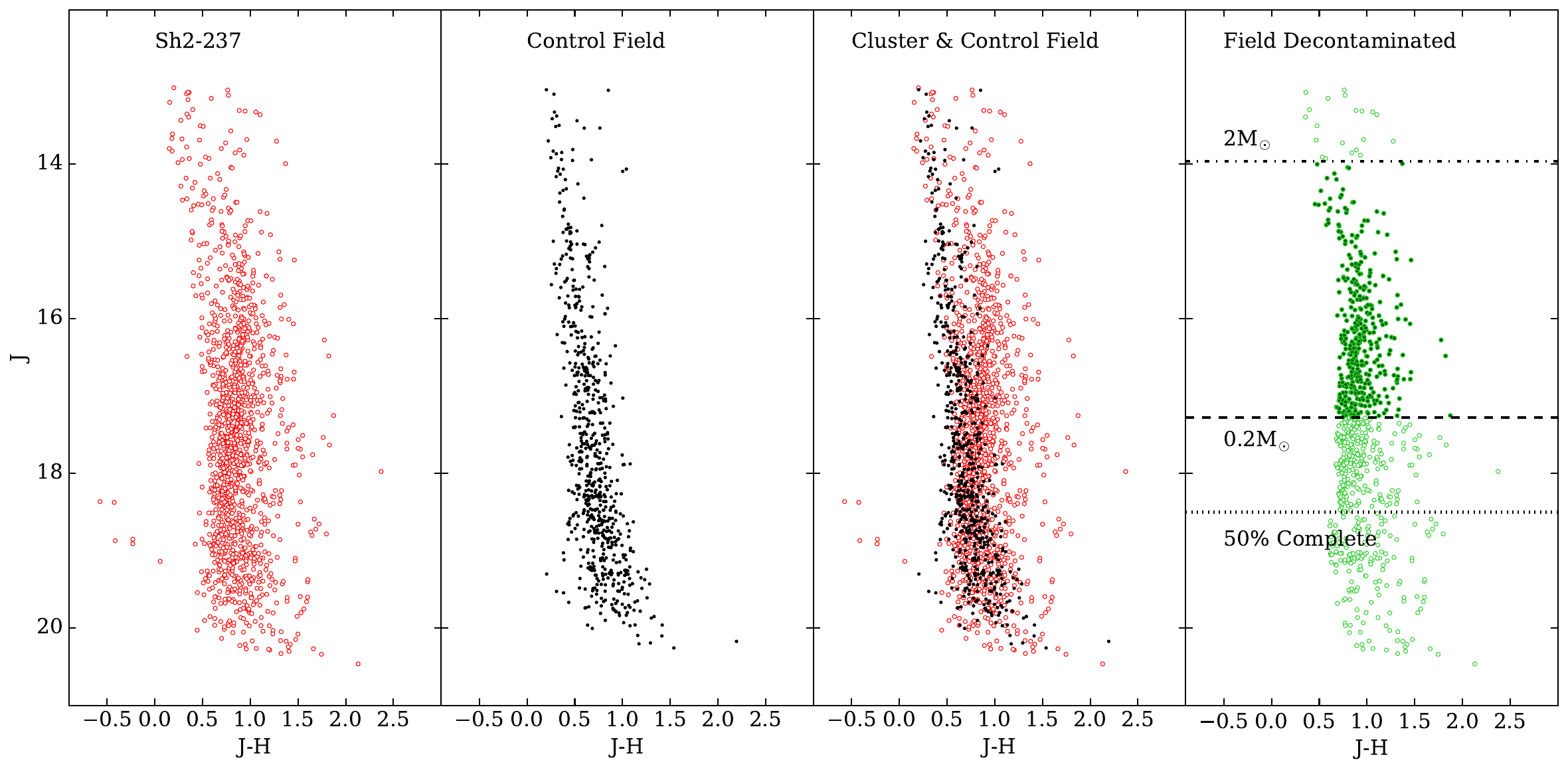}
\caption{CMDs of Sh2-237 \label{Appen_a3}}
\end{figure}

\begin{figure}[ht!]
\epsscale{1}
\plotone{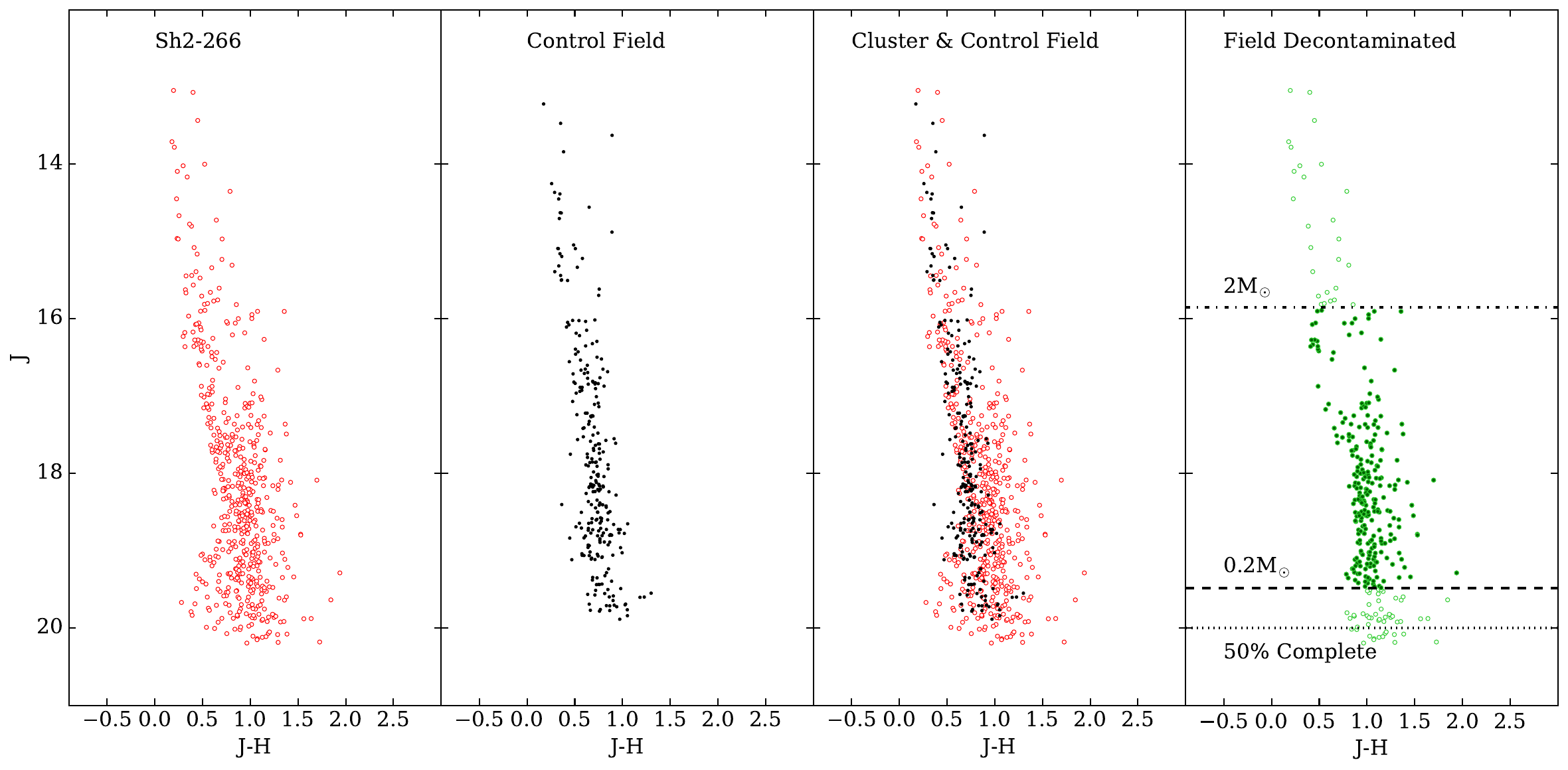}
\caption{CMDs of Sh2-266 \label{Appen_a4}}
\end{figure}

\begin{figure}[ht!]
\epsscale{1}
\plotone{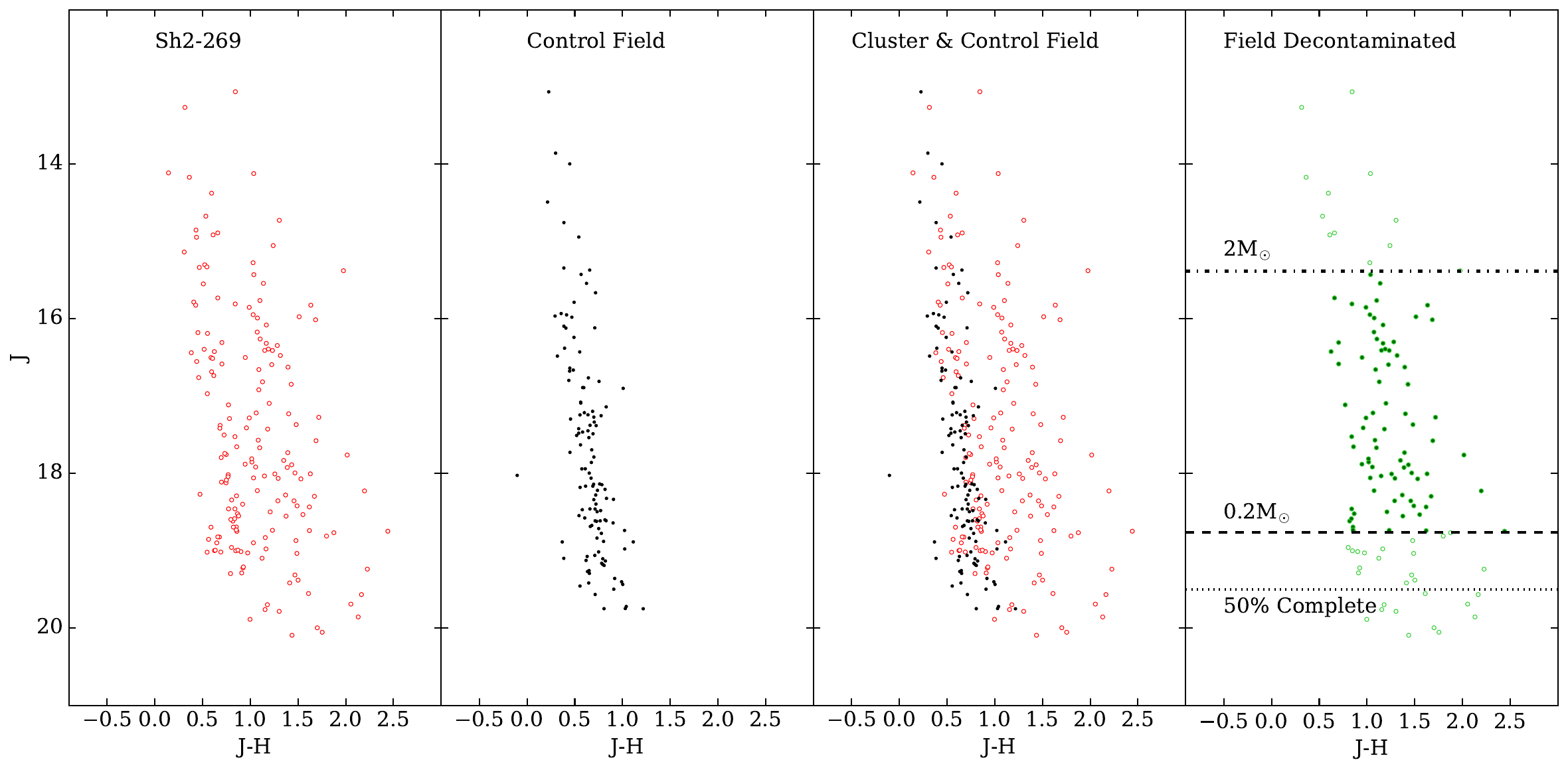}
\caption{CMDs of Sh2-269 \label{Appen_a5}}
\end{figure}

\begin{figure}[ht!]
\epsscale{1}
\plotone{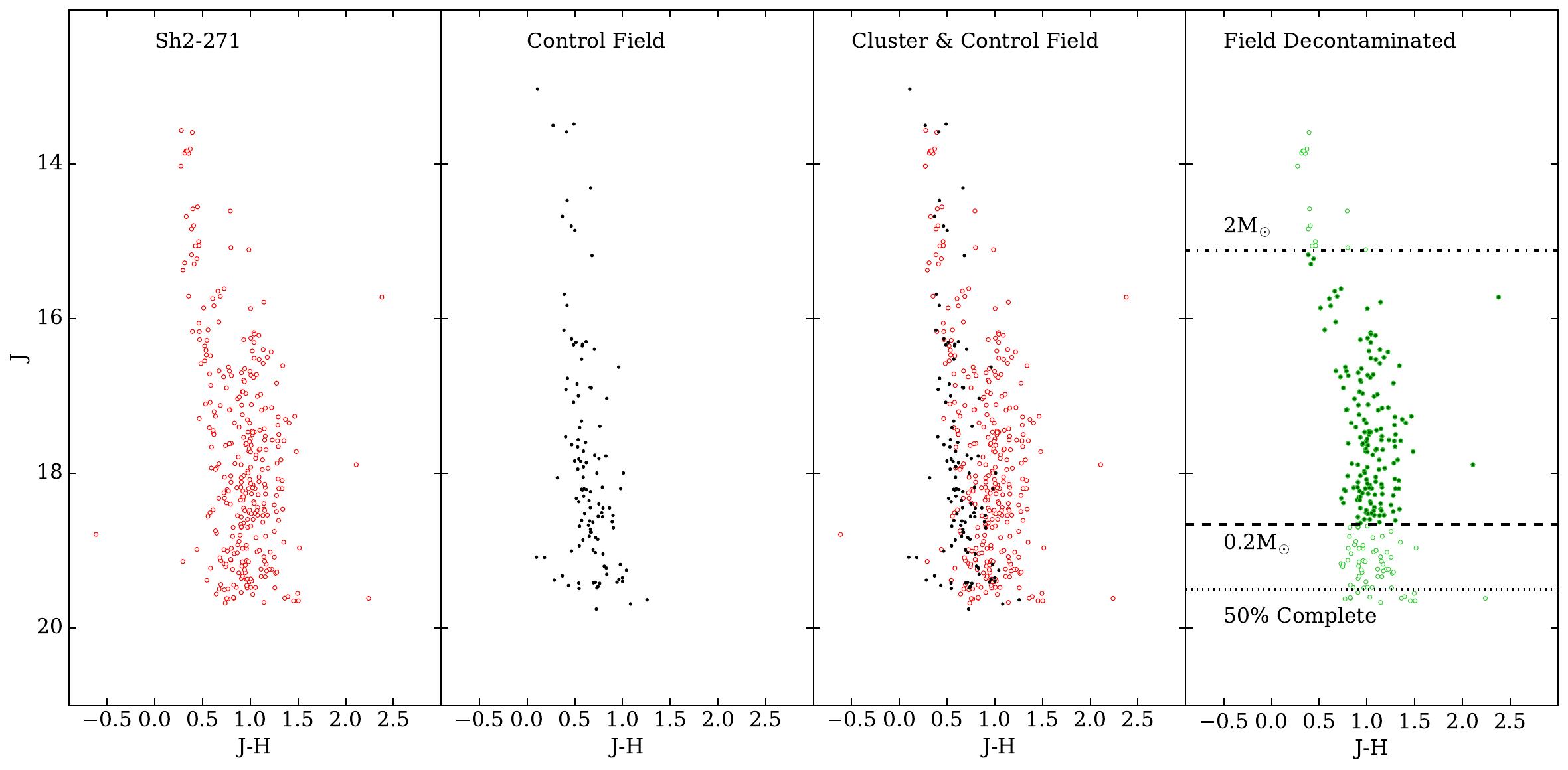}
\caption{CMDs of Sh2-271 \label{Appen_a6}}
\end{figure}

\begin{figure}[ht!]
\epsscale{1}
\plotone{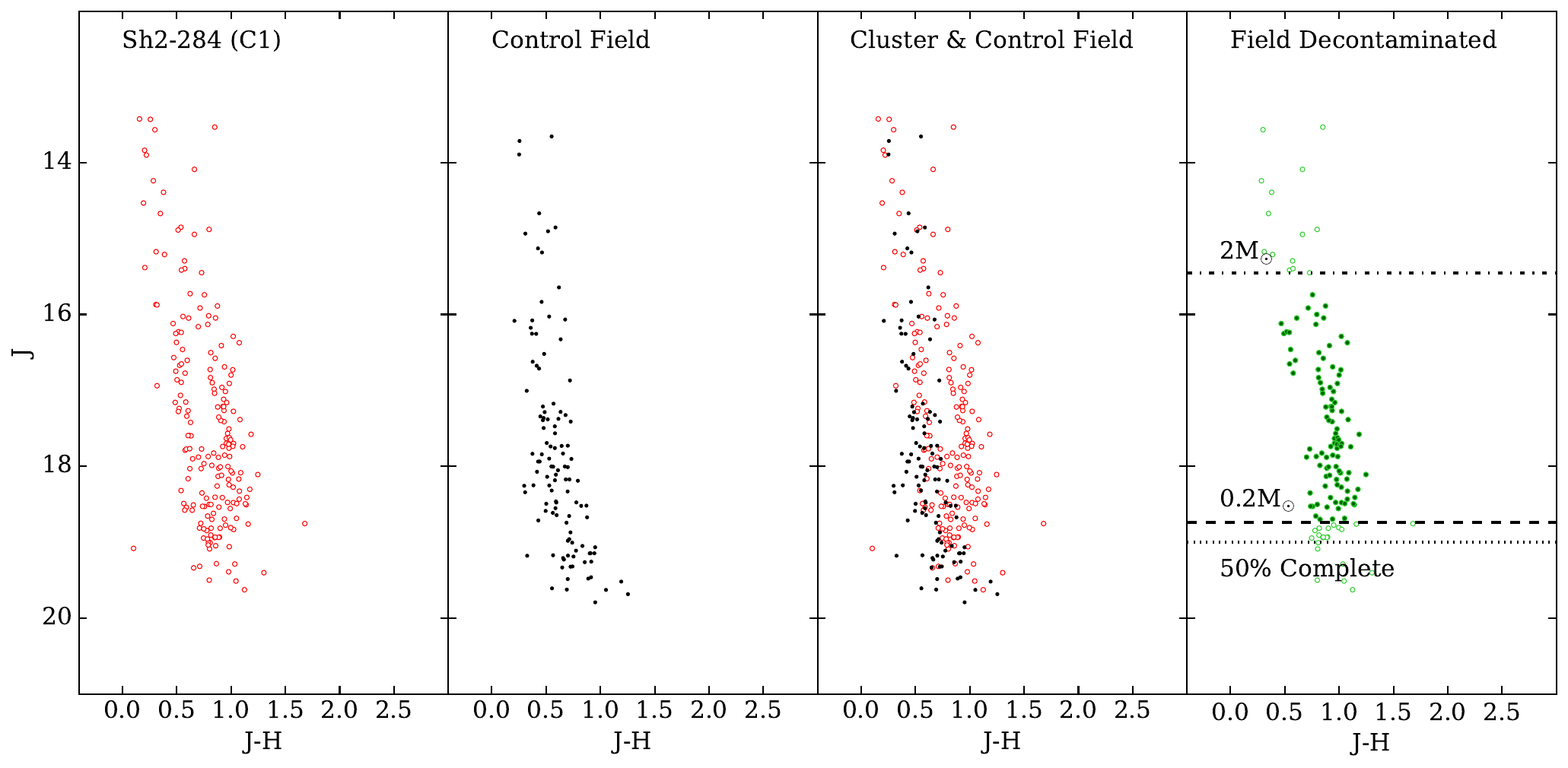}
\caption{CMDs of Sh2-284 (C1) \label{Appen_a7}}
\end{figure}

\begin{figure}[ht!]
\epsscale{1}
\plotone{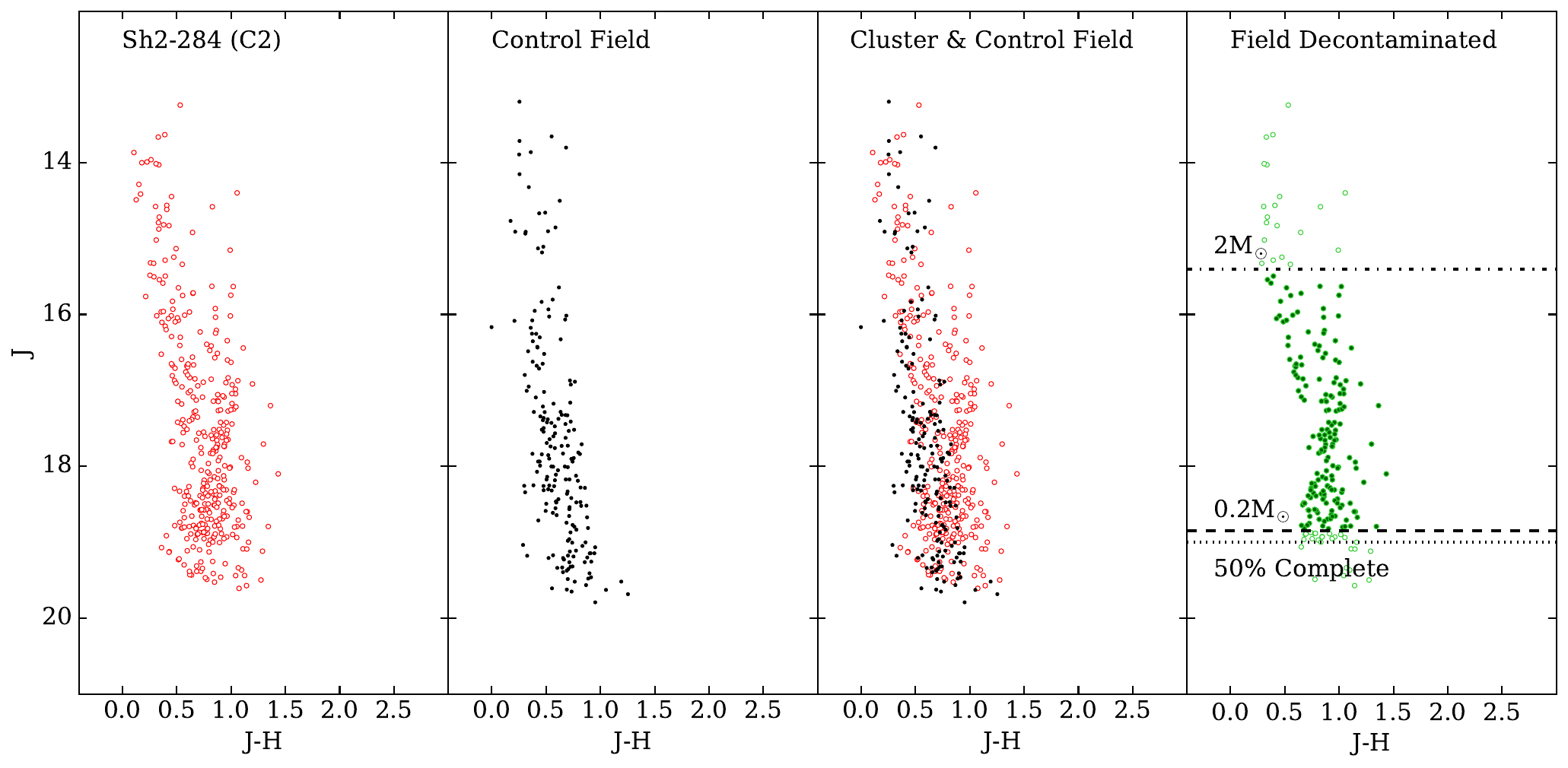}
\caption{CMDs of Sh2-284 (C2) \label{Appen_a8}}
\end{figure}

\begin{figure}[ht!]
\epsscale{1}
\plotone{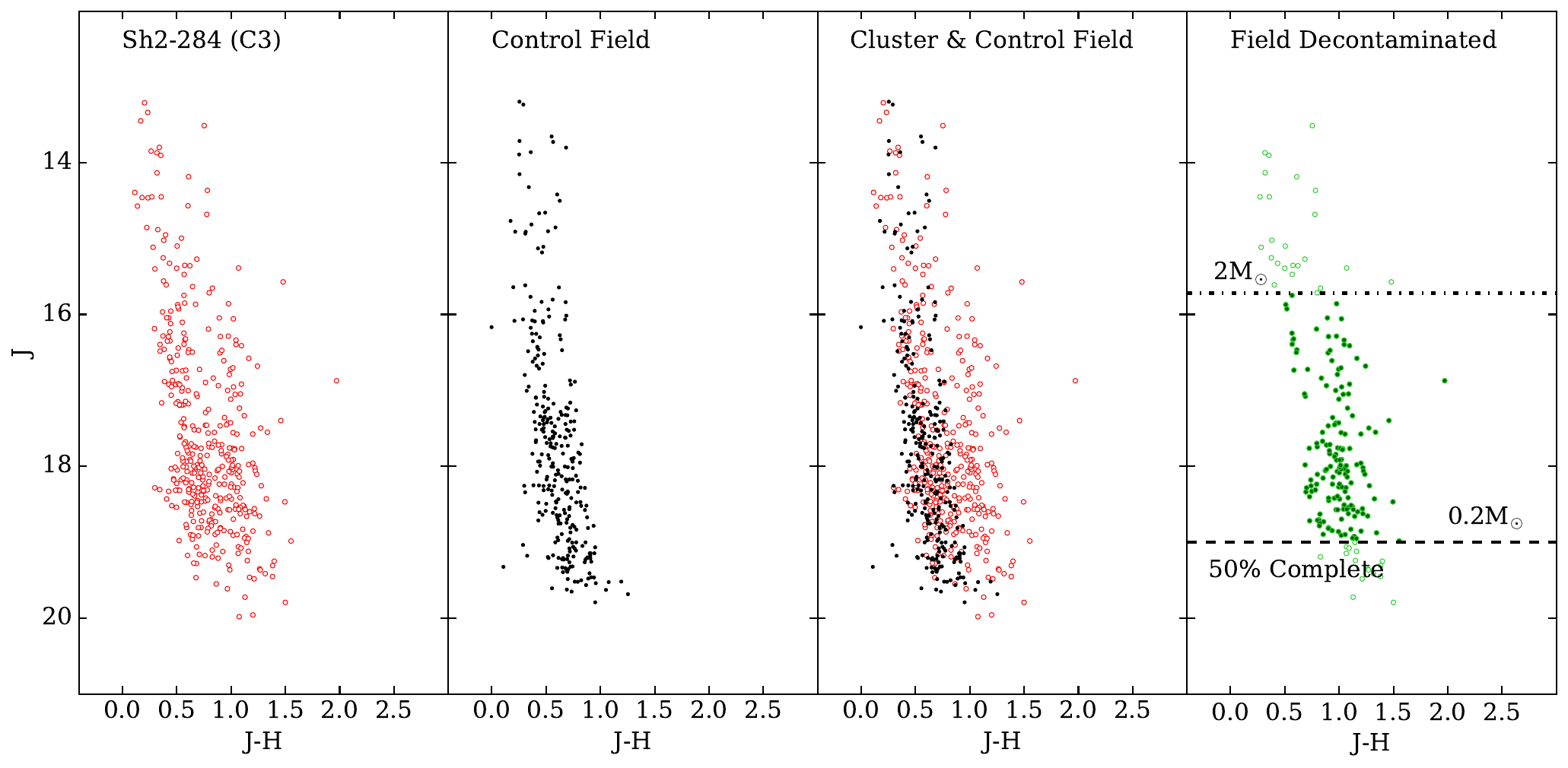}
\caption{CMDs of Sh2-284 (C3) \label{Appen_a9}}
\end{figure}

Figures \ref{Appen_a1}-\ref{Appen_a9} show the representation of the field star decontamination procedure using the color-magnitude diagram (CMD) for Sh2-219, Sh2-228, Sh2-237, Sh2-266, Sh2-269, Sh2-271 and 3 sub-clusters in Sh2-284.

\clearpage
\section{Appendix B}\label{appen:B}
The color-color diagrams are shown in Figure \ref{Appenb:cc_diag1}-\ref{Appenb:cc_diag2} for the same clusters mentioned in Appendix \ref{appen:A}.

\begin{figure*}[h]
\gridline{\fig{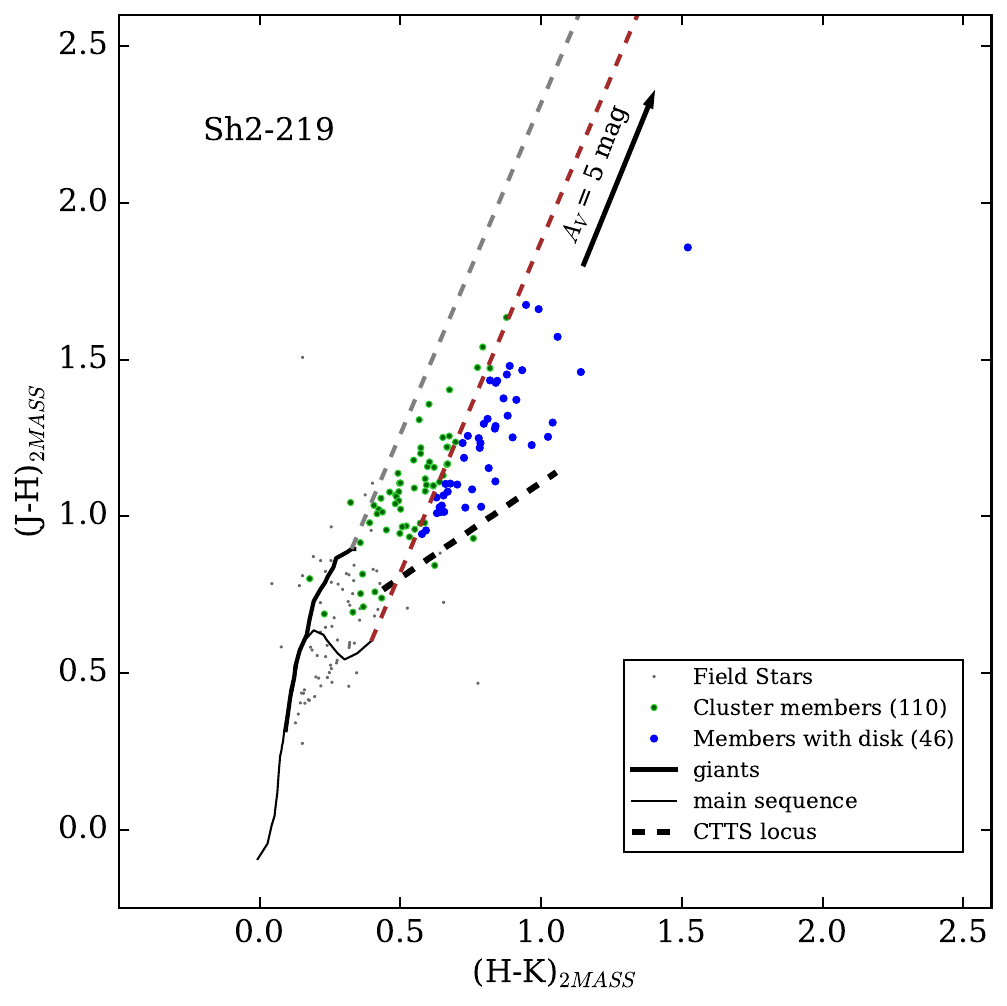}{0.4\textwidth}{(a)}
         \fig{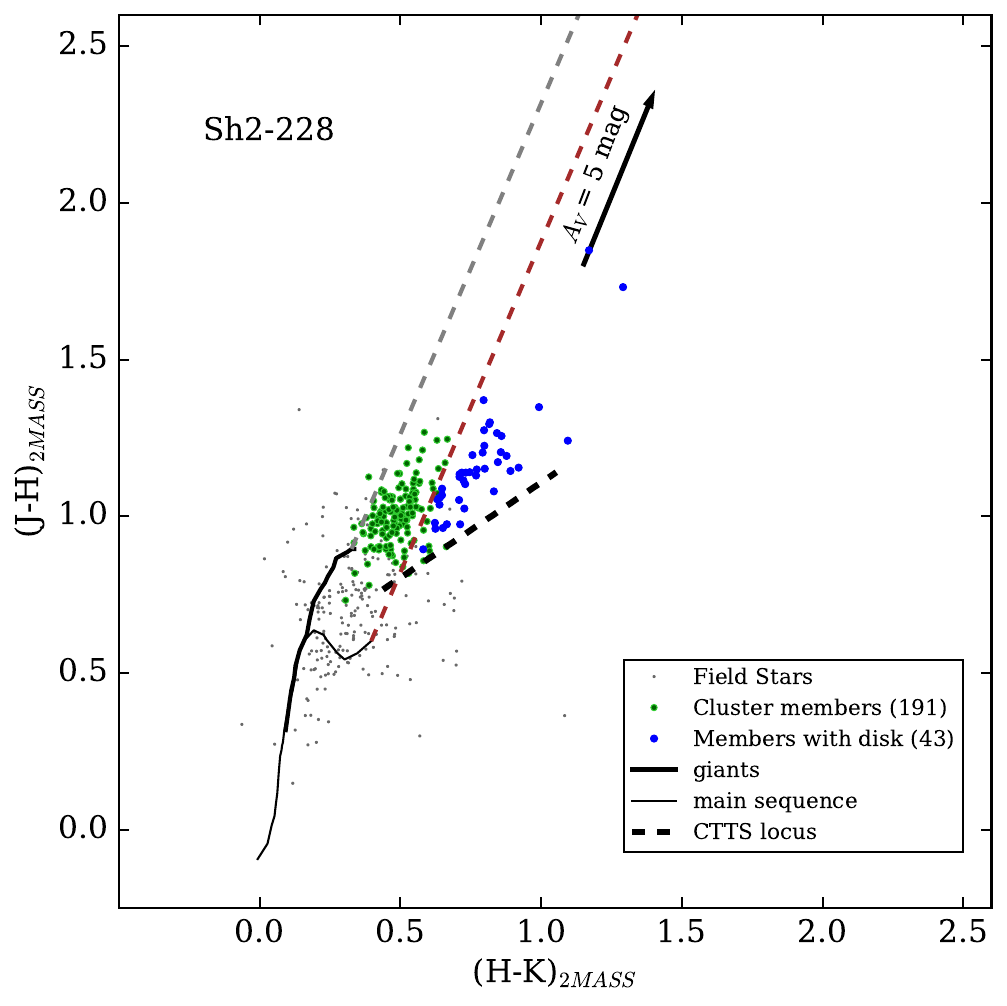}{0.4\textwidth}{(b)}}
\gridline{\fig{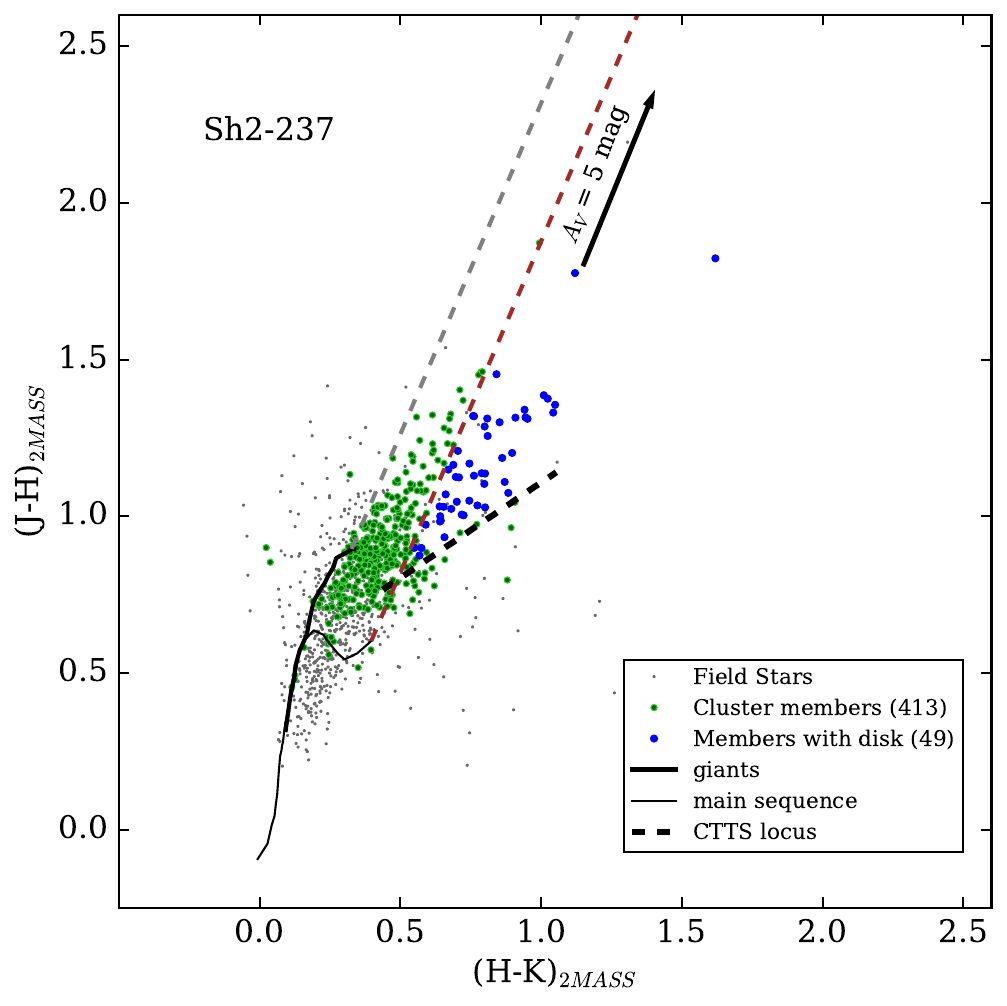}{0.4\textwidth}{(c)}
         \fig{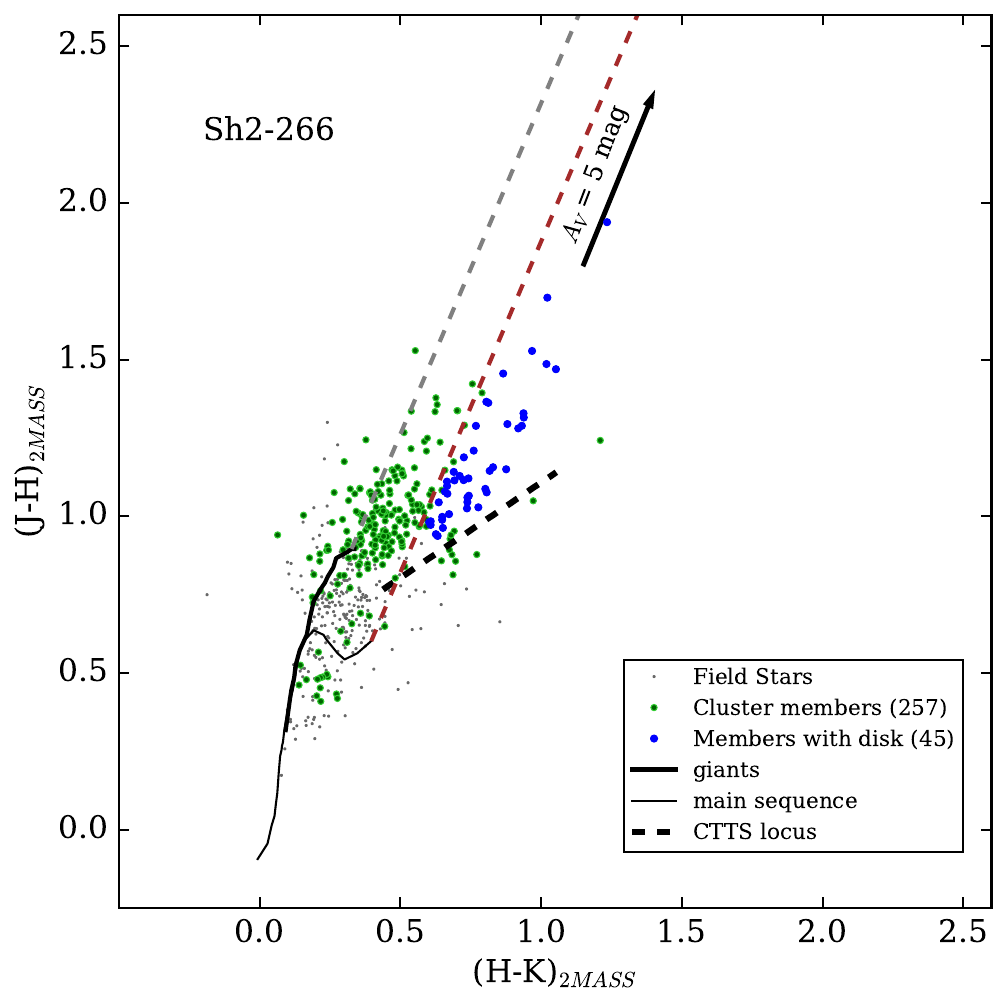}{0.4\textwidth}{(d)}}   
\caption{$(J-H)$ vs. $(H-K)$ color–color diagrams for (a) Sh2-219, (b) Sh2-228, (c) Sh2-237 and (d) Sh2-266 in the 2MASS system.
\label{Appenb:cc_diag1}}
\end{figure*}

\begin{figure*}[h]
\gridline{\fig{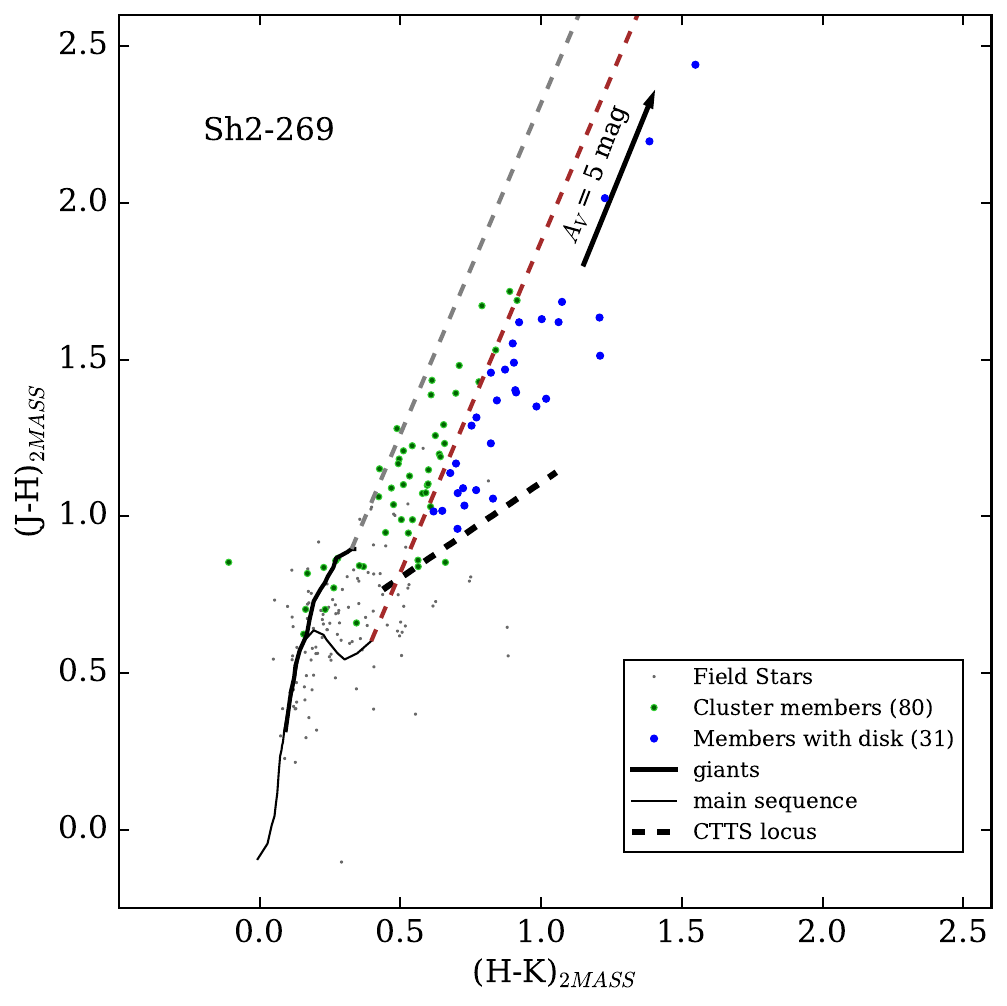}{0.4\textwidth}{(a)}
          \fig{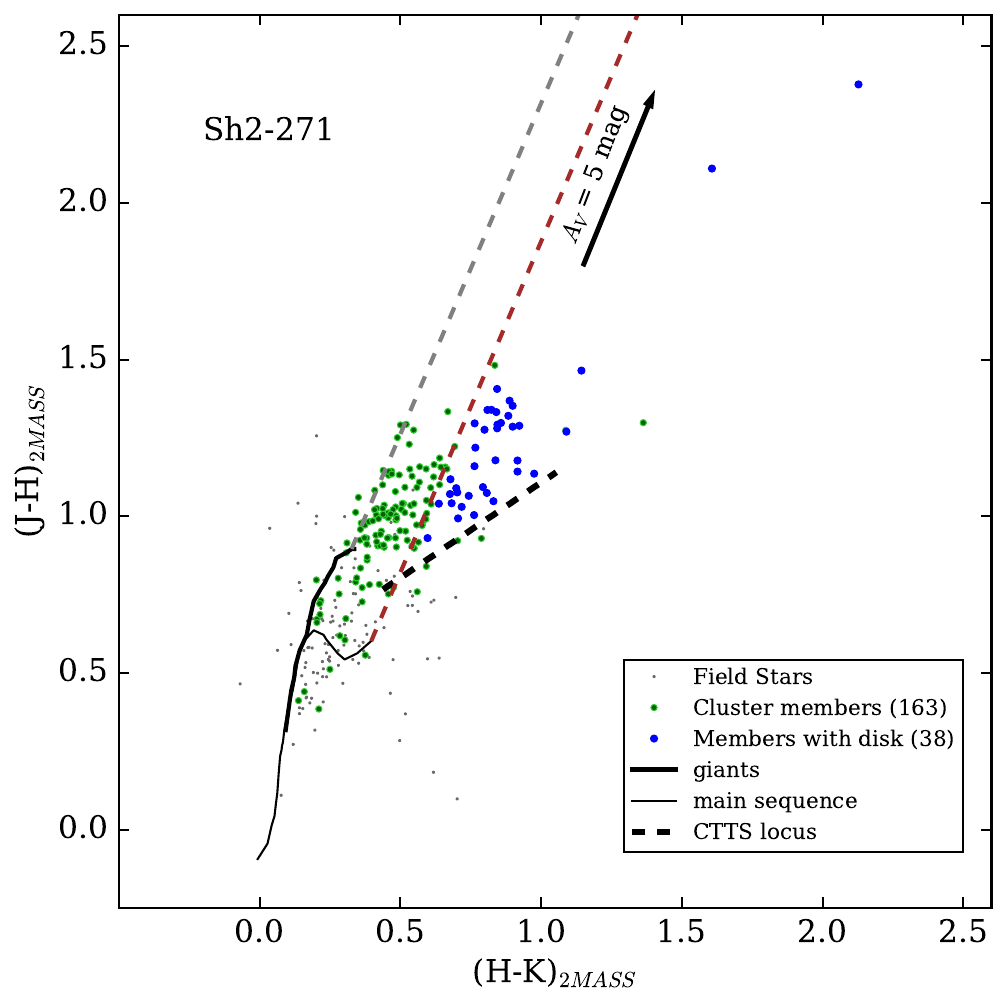}{0.4\textwidth}{(b)}} 
\gridline{\fig{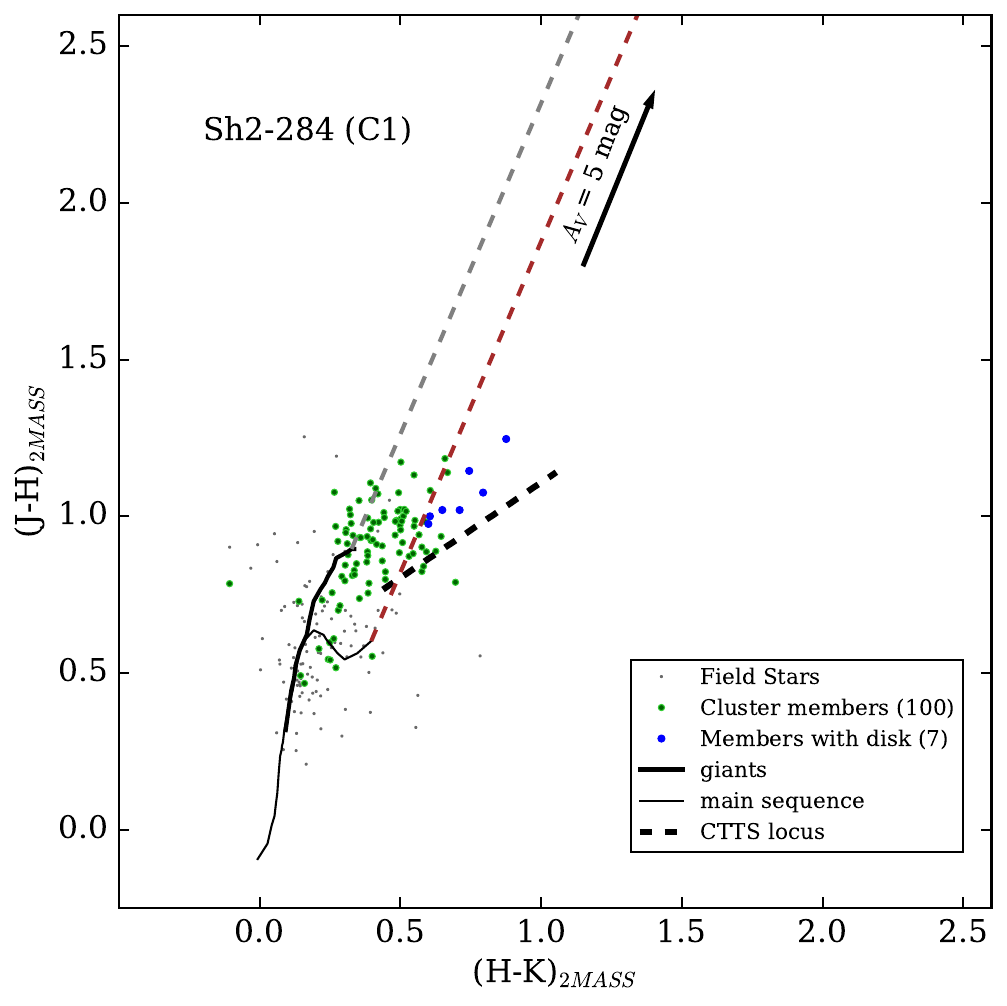}{0.4\textwidth}{(c)}
         \fig{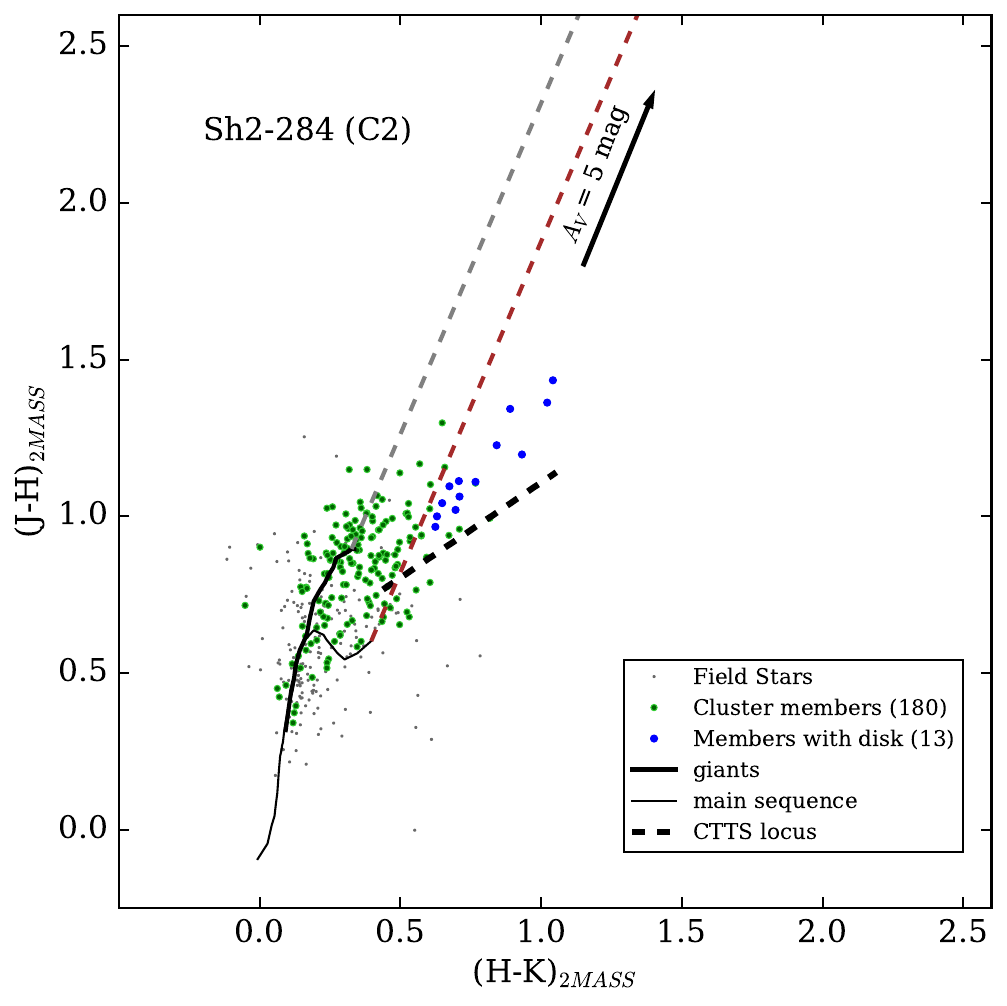}{0.4\textwidth}{(d)}}
\gridline{\fig{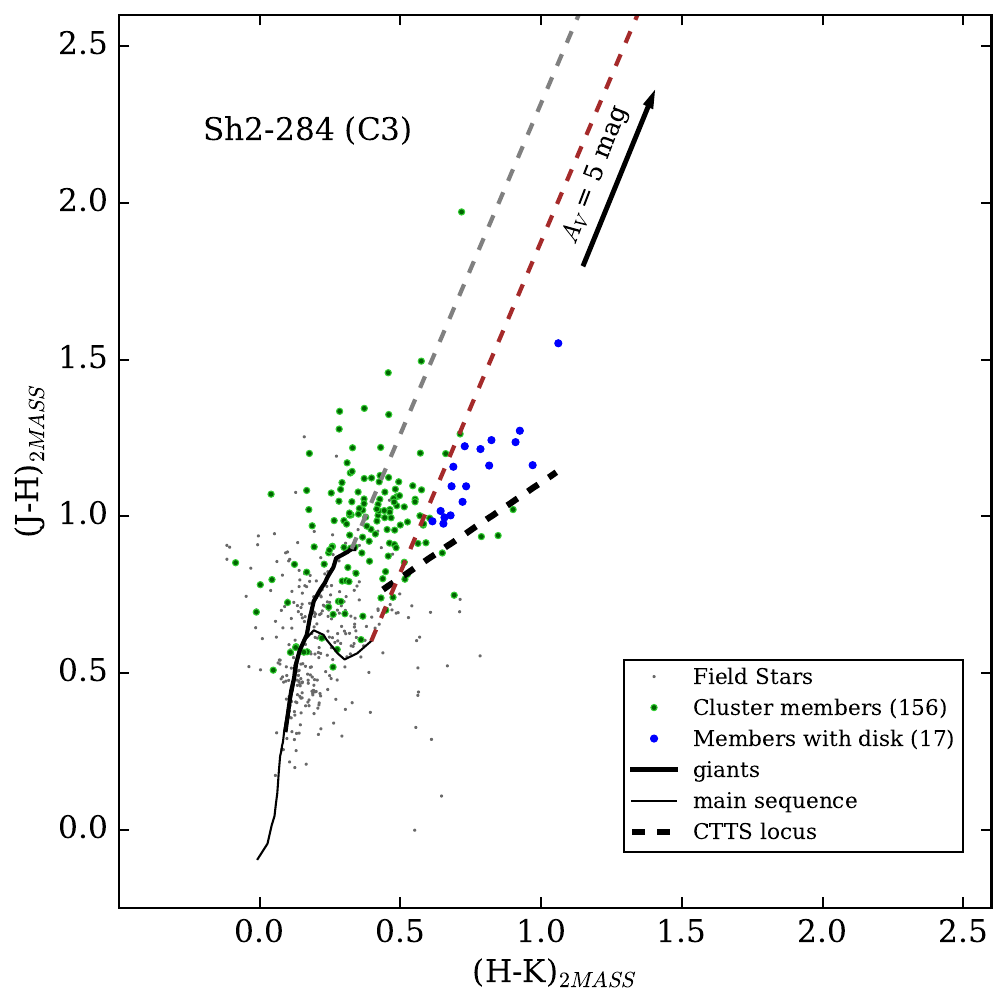}{0.4\textwidth}{(e)}
         }   
\caption{$(J-H)$ vs. $(H-K)$ color–color diagrams for (a) Sh2-269, (b) Sh2-271, (c) Sh2-284 (C1), (d) Sh2-284 (C2) and (e) Sh2-284 (C3) in the 2MASS system.
\label{Appenb:cc_diag2}}
\end{figure*}

\clearpage
\section{Appendix C}\label{appen:c}

\begin{figure}[h!]
\epsscale{0.6}
\plotone{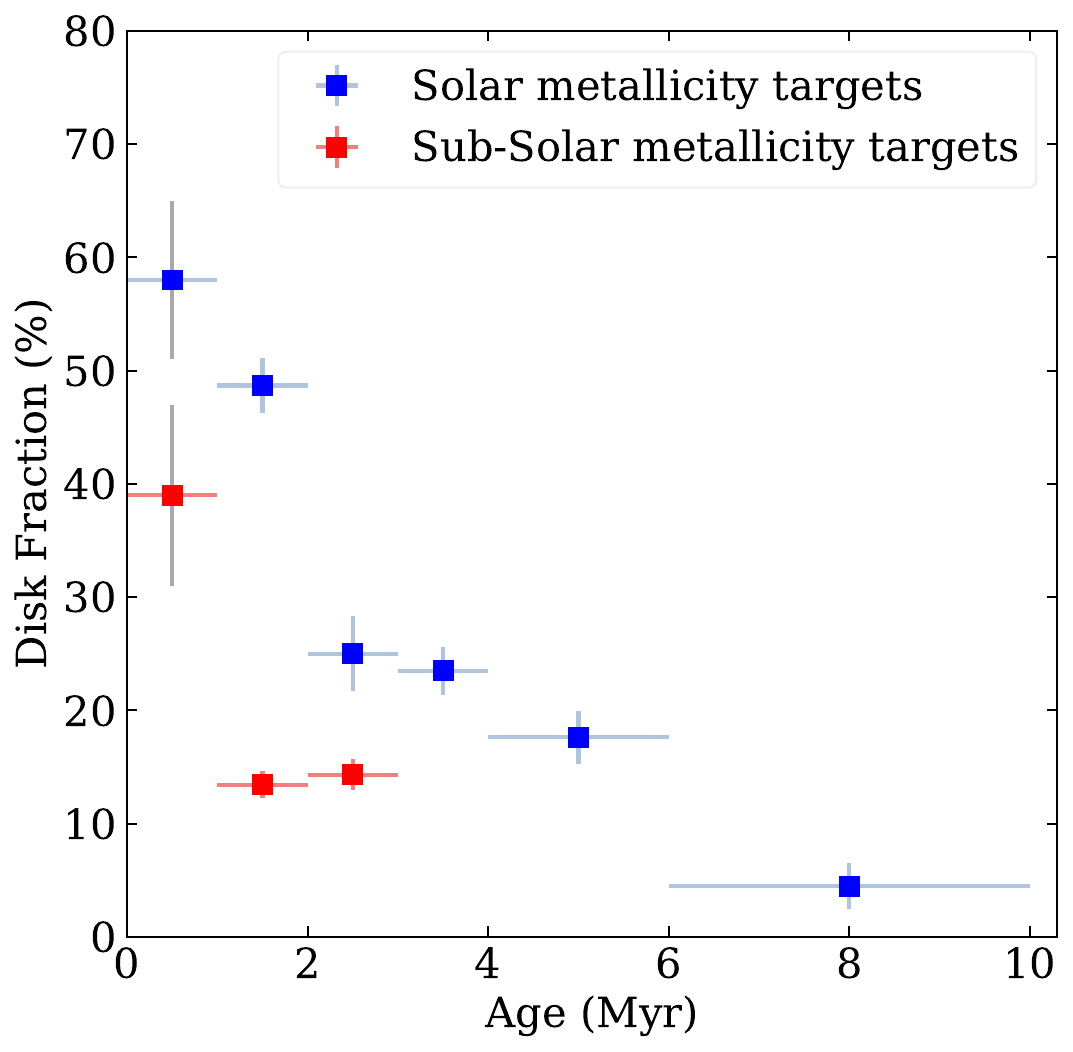}
\caption{Disk fraction ($D_{f}$) vs cluster age ($t$) distribution based on $JHK$ data for the low-metallicity clusters of this work (10 clusters) and solar metallicity regions (12 clusters).  
The solar metallicity clusters (blue squares) are  binned in 1 Myr age span upto 4 Myr and next to points show 2 Myr age binning. The 10 low metallicity clusters studied in this paper are binned in 1 Myr age span upto 3 Myr, denoted by red squares.
\label{fig:df_vs_age_patra}}
\end{figure}

In Figure \ref{fig:df_vs_age_patra}, we present a plot similar to Figure \ref{fig:df_vs_age}, considering only the new 10 low-metallicity targets from this work. The red squares represent the weighted mean of the disk fraction values within each age bin for 10 sub-solar regions. Vertical error bars indicate the errors in the weighted mean of the disk fraction within each age bin, while horizontal error bars depict the age bin spread. For the first bin, where only one target is available, the vertical error bar denotes the individual uncertainty of that target. Similarly, the blue squares represent the average disk fraction for the solar metallicity regions.
Figure \ref{fig:df_vs_age_patra} also shows a trend of lower disk fractions in sub-solar metallicity regions across all age bins, mirroring the findings in Figure \ref{fig:df_vs_age}.
The ratio between the weighted mean of the disk fraction values of solar and sub-solar regions are $1.5\pm0.3$, $3.6\pm0.4$, and $1.7\pm0.3$ for age bins 0-1 Myr, 1-2 Myr, and 2-3 Myr, respectively.

\bibliography{ms}{}
\bibliographystyle{aasjournal}



\end{document}